\documentclass[sigconf]{acmart}

\AtBeginDocument{%
  \providecommand\BibTeX{{%
    \normalfont B\kern-0.5em{\scshape i\kern-0.25em b}\kern-0.8em\TeX}}}

\copyrightyear{2026}
\acmYear{2026}
\setcopyright{cc}
\setcctype{by-nc-nd}
\acmConference[CHI '26]{Proceedings of the 2026 CHI Conference on Human Factors in Computing Systems}{April 13--17, 2026}{Barcelona, Spain}
\acmBooktitle{Proceedings of the 2026 CHI Conference on Human Factors in Computing Systems (CHI '26), April 13--17, 2026, Barcelona, Spain}
\acmPrice{}
\acmDOI{10.1145/3772318.3790374}
\acmISBN{979-8-4007-2278-3/2026/04}


\keywords{AI for Social Good; Funding; Document Analysis; AI imaginaries}

\usepackage{graphicx} 
\usepackage{xcolor} 
\usepackage{lscape} 
\usepackage{longtable} 
\usepackage{caption}   
\usepackage{float}
\usepackage{hyperref}

\begin{document}

\newcommand{\idanchor}[1]{\hypertarget{id:#1}{#1}}
\newcommand{\id}[1]{\hyperlink{id:#1}{#1}}

\newcommand\anna[1]{\textcolor{red}{AK: #1}}
\newcommand\ken[1]{\textcolor{magenta}{KH: #1}}

\title{Funding AI for Good: A Call for Meaningful Engagement}

\author{Hongjin Lin}
\orcid{0000-0001-6207-2147}
\affiliation{%
  \department{School of Engineering and Applied Sciences}
  \institution{Harvard University}
  \city{Cambridge}
  \state{Massachusetts}
  \country{USA}}
\email{hongjin_lin@g.harvard.edu}

\author{Anna Kawakami}
\orcid{0009-0000-6937-6370}
\affiliation{%
  \department{Human-Computer Interaction Institute}
  \institution{Carnegie Mellon University}
  \city{Pittsburgh}
  \state{Pennsylvania}
  \country{USA}}
\email{akawakam@andrew.cmu.edu}

\author{Catherine D’Ignazio}
\orcid{0000-0002-8673-1941}
\affiliation{%
  \department{Dept of Urban Studies \& Planning}
  \institution{MIT}
  \city{Cambridge}
  \state{Massachusetts}
  \country{USA}}
\email{dignazio@mit.edu}

\author{Kenneth Holstein}
\orcid{0000-0001-6730-922X}
\affiliation{%
  \department{Human-Computer Interaction Institute}
  \institution{Carnegie Mellon University}
  \city{Pittsburgh}
  \state{Pennsylvania}
  \country{USA}}
\email{kjholste@cs.cmu.edu}

\author{Krzysztof Z. Gajos}
\orcid{0000-0002-1897-9048}
\affiliation{%
  \department{School of Engineering and Applied Sciences}
  \institution{Harvard University}
  \city{Cambridge}
  \state{Massachusetts}
  \country{USA}}
\email{kgajos@g.harvard.edu}

\renewcommand{\shortauthors}{Lin et al.}

\newcommand{\hongjin}[1]{{\color{cyan} Hongjin: #1}}

\begin{abstract}
Artificial Intelligence for Social Good (AI4SG) is a growing area that explores AI's potential to address social issues, such as public health. Yet prior work has shown limited evidence of its tangible benefits for intended communities, and projects frequently face real-world deployment and sustainability challenges. While existing HCI literature on AI4SG initiatives primarily focuses on the mechanisms of \textit{funded projects} and their outcomes, much less attention has been given to the upstream \textit{funding agendas} that influence project approaches. In this work, we conducted a reflexive thematic analysis of 35 funding documents, representing about \$410 million USD in total investments. We uncovered a spectrum of conceptual framings of AI4SG and the approaches that funding rhetoric promoted: from biasing towards technology capacities (more \textit{techno-centric}) to emphasizing contextual understanding of the social problems at hand alongside technology capacities (more \textit{balanced}). Drawing on our findings on how funding documents construct AI4SG, we offer recommendations for funders to embed more \textit{balanced} approaches in future funding call designs. We further discuss implications for how the HCI community can positively shape AI4SG funding design processes.

\end{abstract}

\maketitle

\section{Introduction}

\begin{figure*}
    \centering
    \includegraphics[width=\linewidth]{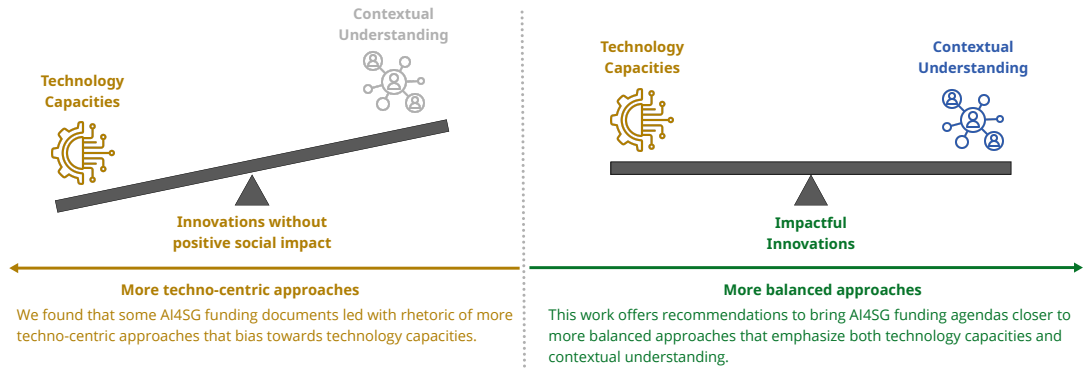}
    \caption{AI4SG aims to bring about positive, long-lasting social impact. Prior literature emphasizes that impactful innovations are the result of a balance of technology capacities and contextual understanding (right panel). A bias towards technology capacities leads to innovations without a real-world positive impact (left panel). We use this spectrum as a lens to investigate how current AI4SG funding documents orient towards more \textit{techno-centric} or \textit{balanced} approaches.}
    \Description{This figure shows a spectrum of two approaches in pursuing technology development for social impact. The left panel shows an overemphasis on technical problems (a more techno-centric approach), which could lead to innovations without social impact. Our work found that some AI4SG funding documents led with rhetoric of more techno-centric approaches that are biased towards technological capacities. The right panel shows a balance of emphasis on technical capacities and contextual understanding (a more balanced approach), which could lead to impactful innovations. This work offers recommendations to bring AI4SG funding agendas closer to more balanced approaches that emphasize both technology capacities and contextual understanding.}
    \label{fig:framing}
\end{figure*}


Artificial Intelligence (AI) for social good (AI4SG) initiatives aim to create long-lasting positive impact in social domains such as public health, by leveraging AI technologies like machine learning~\cite{shi_artificial_2020, li_ai_2021, perrault_ai_2020, rolnick_tackling_2022, tomasev_ai_2020}. Yet prior work finds limited evidence of tangible benefits of these initiatives for intended communities, and projects frequently face deployment and sustainability challenges in real-world contexts~\cite{vinuesa_role_2020, eubanks_automating_2018, sally_ho_heres_2023, tate_ryan-mosley_algorithm_2023}. HCI scholarship provides insights into the socio-technical mechanisms underpinning these challenges, and recent work points to the significant influence of funding agendas---the priorities, goals, and approaches set by funders---in shaping the technology initiatives aimed at social impact~\cite{lin_come_2024, saha_commissioning_2022, erete_method_2023}. In AI4SG specifically, \citet{lin_come_2024} foreground community organizations' perspectives and show that funding agendas heavily influence how partnerships are formed, which problem spaces are prioritized, and what kinds of solutions are pursued during the early stages of projects. Prior work also notes that funding often incentivizes mission-driven organizations to adopt data-driven innovations, sometimes at the cost of their original social missions~\cite{bopp_disempowered_2017, lin_come_2024}. For AI developers and practitioners, short funding timelines can constrain considerations of ethics and unintended consequences~\cite{do_thats_2023, moitra_ai_2022, okolo_you_2024}. However, existing HCI literature on AI4SG initiatives primarily focuses on \textit{funded projects}' approaches and outcomes through the perspectives of project teams; much less attention is given to the design of upstream \textit{funding agendas} themselves, before projects even get selected.

In this work, we begin addressing this gap by investigating AI4SG funding agendas, following up on Laura Nader's call to ``study up''~\cite{nader_up_1972} systems and actors that exert normative influence. In doing so, we aim to shift attention from assigning blame to AI4SG project teams (typically consisting of AI technologists and mission-driven organizations), and towards recognizing the broader power structures shaping their approaches. We turn to public funding documents---textual instruments that construct and communicate funding agendas, including calls for proposals and grant announcements. Prior literature has established that funding documents are not neutral instruments: they construct priorities~\cite{palmero_thematic_2025}, shape applicant behaviors~\cite{peng_promotional_2024}, and structure downstream research activities~\cite{widder_basic_2024, smith_understanding_2023}. In AI4SG, by defining \textit{who} is eligible, \textit{what} constitutes a problem, \textit{how} these problems should be addressed, and \textit{what} success means, funding documents construct discourses about ``AI'' and ``social good'', signaling what approaches are fundable or imaginable. Although the mechanisms by which written funding documents influence what actually gets funded are complex, examining funding rhetoric reveals how priorities, values, and power relations are embedded before any project begins. Funding documents can reinforce dominant \textit{AI imaginaries}---``collective visions, beliefs, symbols, and expectations that individuals and communities hold about AI and its potential outcomes''~\cite{zhong_ai_2025}---as seen in public texts portraying AI as ``inevitable and massively disrupting''~\cite{bareis_talking_2022}. Yet they can also guide applicants toward community engagement strategies previously unfamiliar to them.

In analyzing funding documents, our goal is not to evaluate how they eventually lead to funded project outcomes, but to \textbf{provide thematic insights into the rhetoric and conceptual framing of AI4SG in funding documents that influence downstream approaches}. Specifically, we ask the following research questions:

\begin{itemize}
    \item \textbf{RQ1}: How do funding documents frame social good and AI's role in advancing social good?
    \item \textbf{RQ2}: How do funding documents describe the expected outcomes of AI4SG projects?
    \item \textbf{RQ3}: How do funding documents envision how these goals might be achieved?
\end{itemize}

We collected and conducted a reflexive thematic analysis of 35 funding documents---19 funding calls and 16 grant announcements---representing about \$410 million USD in total investments (see Appendix~\ref{app:funding_doc} for a complete list). These documents represent a range of efforts, broadly framed under the banners of AI or data-driven technologies for ``social good,'' ``development,'' and other social domains. Our dataset includes funding programs from 23 distinct funders, including private technology companies, philanthropic foundations, nonprofit organizations, and government agencies based in Western societies such as the United States.

Since AI4SG builds on earlier ``tech for good'' efforts~\cite{aula_stepping_2023}, we ground our analysis in lessons learned from Computer-Supported Cooperative Work (CSCW), Information and Communication Technologies and Development (ICTD), and Human-Computer Interaction for Development (HCI4D). This prior literature has extensively examined different approaches in socially driven technology projects, emphasizing that impactful innovations require both a thorough understanding of the problem and appropriate technological capabilities (Figure~\ref{fig:framing}). From our synthesis of this body of work, a spectrum from \textit{techno-centric} to \textit{balanced} approaches emerges, indicating a strong association between where projects fall on this spectrum and project outcomes. A \textit{techno-centric} approach assumes novel technology alone can solve complex social issues, often producing suboptimal results and strained relationships between researchers and communities (e.g.,~\cite{ames_charisma_2019, karusala_womens_2017, walsham_ict4d_2017, dada_failure_2006, brown_why_2019}). In contrast, a more \textit{balanced} approach, grounded in a solid understanding of the community's challenges and contexts through community-collaborative approaches~\cite{cooper_fitting_2024}, tend to result in more community buy-in and sustainability (e.g.,~\cite{toyama_geek_2015, gandhi_digital_2007, burrell_what_2009}). Using this spectrum, we examine how current AI4SG funding rhetoric is oriented towards a more \textit{techno-centric} or a more \textit{balanced} approach. 

Funding documents in our corpus overwhelmingly mobilized AI against large-scale social problems that cannot be fully addressed with any single intervention (notably public health and development). Yet several funding documents framed AI's potential as deterministically beneficial and positioned AI interventions as solutions to these social problems while skirting contextual and stakeholder specificity. Some other documents scoped interventions to more well-defined sub-problems and explicitly acknowledged both AI's benefits and risks. Expected outcomes expressed in the funding documents in our dataset frequently centered on technical deliverables, while a subset adopted broader success metrics attentive to benefits to the communities involved. Furthermore, although interdisciplinary collaboration was widely encouraged, funding documents varied in their emphasis and scaffolding of community engagement and who were considered ``experts.'' With a few exceptions, expectations for community engagement stopped at a consultation mode towards later stages of projects (e.g., implementation stage), where community stakeholders were expected to provide inputs but not be able to shape the broader project scopes.

To the best of our knowledge, this study presents the first qualitative analysis of how AI4SG funding documents frame social good, envision AI’s role in advancing it, and anticipate desired outcomes to be achieved. While prior work has largely focused on funded projects' tendency towards more \textit{techno-centric} approaches, we uncover \textit{how} both \textit{techno-centric} and \textit{balanced} approaches are instantiated in different components of funding documents---before project selection even occurs. In doing so, this work makes three contributions to existing HCI literature on ``tech for good'' initiatives and document analysis of AI imaginaries: 
\begin{itemize}
    \item First, we provide a novel focus on AI4SG funding documents as a discourse instrument rather than neutral texts. Through a reflexive thematic analysis~\cite{braun_using_2006, braun_reflecting_2019}, we highlight the disconnects between the positive social impact AI4SG aims for and the \textit{techno-centric} approaches promoted in some funding documents (Section~\ref{sec:discussion1}).
    \item Second, building on the more \textit{balanced} approaches and gaps identified in our dataset, we propose opportunities for designing future funding calls for funders (including supporting relationship-building, maintenance post-project deployment, and community engagement training for project teams), and outline future research directions for HCI researchers (Section~\ref{sec:discussion2}).
    \item Finally, we call for meaningful community engagement in the funding design process for funders, and specify the roles that HCI practitioners and researchers can play in shaping AI4SG funding agendas and design processes (Section~\ref{sec:discussion3}).
\end{itemize}

\section{Related Work}

We draw on prior HCI research on AI4SG, which has highlighted the objectives of existing funded projects to produce positive social impact, the challenges they face in achieving them, and the central role of funding agendas in shaping the AI4SG ecosystem (Section~\ref{sec:rw_1}). Scholars have called on AI4SG to draw on lessons from previous ``tech for good'' projects, which emphasize that impactful innovations require a balance of appropriate technical capacity and deep contextual understanding (Section~\ref{sec:lens}). We then draw on emerging scholarship examining grant-making and private tech companies' investment in AI4SG (Section~\ref{sec:rw_3}). Lastly, we draw on prior research in HCI and innovation studies that use document analysis to examine how AI is imagined in public texts, and emphasize the role of funding documents in shaping research priorities and collaboration structures (Section~\ref{document_analysis}). 

HCI has yet to systematically analyze how funding documents discursively define and construct AI4SG initiatives and the conditions under which their positive impact might be achieved. This study begins to address this gap by analyzing AI4SG funding documents through the lens of existing literature on ``tech for good.'' While prior work points out funded projects' bias towards technical capacity, our work contributes an analysis of \textit{how} this bias is instantiated across some funding documents, while also identifying more balanced approaches that emphasize contextual understanding.

\subsection{The Promises and Challenges of AI for Social Good}
\label{sec:rw_1}

Artificial Intelligence for Social Good (AI4SG) has emerged as a growing field of use-inspired research and real-world implementations focused on applying AI to complex social issues~\cite{shi_artificial_2020, perrault_ai_2020, tomasev_ai_2020, rolnick_tackling_2022}. Its stated goal is to deliver transformative, positive, and lasting impact across many social aspects~\cite{perera_scaling_2024, cowls_ai_2021, tudor_leveraging_2024}. Some notable examples include projects in agriculture that predict crop yields using remote sensing data~\cite{you_deep_2017} and provide price forecasting for farmers in India~\cite{ma_interpretable_2019}. 
While enthusiasm for AI4SG remains high, growing evidence shows that many projects fall short of delivering lasting positive social impact and often end at the publication stage without deployment~\cite{lin_come_2024, aula_stepping_2023, ben_green_good_2019}. In high-stakes contexts such as public service screening or criminal justice, AI may be inappropriate altogether~\cite{eubanks_automating_2018, sally_ho_heres_2023, pruss_ghosting_2023}. Additionally, AI4SG is often shaped by promotional hype~\cite{joyce_sociology_2024}, magical thinking~\cite{lupetti_making_2024}, and ``enchanted determinism''~\cite{campolo_enchanted_2020}, where the fascination with the mysterious capabilities of new technologies captures public attention toward AI interventions. This could be exacerbated by AI's transparency issues (especially in the Global South context that AI4SG projects are intended for~\cite{okolo_you_2024}). For instance, studies in India show a strong belief in ``AI authority'' over public institutions and lowered expectations for accountability~\cite{kapania_because_2022,ramesh_how_2022}. 


HCI scholars have started to examine the mechanisms underlying AI4SG projects. We highlight two key insights below. First, the aspiration to employ a powerful emerging technology for social good is not a recent phenomenon; AI4SG should draw lessons learned from prior ``tech for good'' movements~\cite{aula_stepping_2023, lin_come_2024}. ICTD and HCI4D have long explored the development and use of emerging technologies in marginalized communities, especially in the Global South~\cite{walsham_ict4d_2017, dell_ins_2016, avgerou_discourses_2010}. Second, while past work often focused on investigating perspectives of \textit{project teams}---typically AI developers and nonprofit organizations, emerging research highlights \textit{funding agendas’} role in shaping interdisciplinary partnerships~\cite{lin_come_2024, saha_commissioning_2022, erete_method_2023}. Through interviews with staff from community organizations participating in AI4SG, \citet{lin_come_2024} found that funding agendas influence program goals, proposed solutions, and definitions of success. Focusing on nonprofit organizations, other studies show how funding agendas shape their incentives to adopt emerging tech, and their collaboration choices~\cite{bopp_disempowered_2017, erete_storytelling_2016, goldenfein_tech_2023}. AI developers and researchers' approaches to AI ethics and considerations of unintended consequences are also shown to be impacted by funding timeline~\cite{do_thats_2023, widder_its_2023, moitra_ai_2022}. 

Emphasizing the need to examine how power structures impact technology development in social contexts, scholars have taken up Laura Nader’s call to ``study up'' by focusing on systems and actors that shape social systems and exert normative influence~\cite{nader_up_1972}. Emerging ``study up'' scholarship in algorithmic fairness~\cite{barabas_studying_2020}, public sector AI~\cite{kawakami_studying_2024}, and machine learning data~\cite{miceli_studying_2022} demonstrates that examining those in power can reveal critical leverage points for more accountable AI development. While prior work has highlighted funding's significant influence on project approaches and outcomes, we still lack a direct understanding of how funding frames AI’s role in addressing social good challenges; there is a need to ``study up'' AI4SG funding.

\subsection{Lessons from Prior ``Tech for Good'' Projects: Techno-Centric vs Balanced Approaches}
\label{sec:lens}

Prior work in CSCW, ICTD, and HCI4D has established a substantial body of work on technology projects aimed at social impact. Despite existing areas of debate within this body of work, there is a broad consensus that impactful innovations require not only engineering expertise but also situated knowledge and community engagement. Review studies on ICTD projects specifically point out that a bias towards technical problems while overlooking community contexts often leads to limited adoption and community impact~\cite{brown_why_2019, heeks_most_2003, dodson_considering_2012}, whereas sustained community engagement throughout the project life-cycle is a key factor for success~\cite{brown_why_2019}.

In this section, we synthesize this rich body of work and categorize project characteristics into a more \textit{techno-centric} approach (bias towards technical solutions) and a more \textit{balanced} approach (emphasizing both technical capacities and contextual understanding). 
We particularly focus on aspects of motivation, collaboration, longevity, process, and evaluation because these are closely influenced by funding agendas~\cite{lin_come_2024}. 

\subsubsection{Techno-Centric Approaches} A \textit{techno-centric} approach, as defined by \citet{dodson_considering_2012}, ``place[s] the technology at the center of the intervention by giving prominence to technical features and capabilities.'' It is rooted in ``techno-solutionism,'' where complex, structural challenges are reduced to problems amenable to technical interventions~\cite{morozov_save_2013}, often under the false assumption that these solutions are neutral or apolitical~\cite{feenberg_transforming_2002, winner_artifacts_1980}. While leveraging emerging technology as a silver bullet to social problems appears optimistic and well-intentioned, Berlant’s concept of ``cruel optimism'' points to the dark side of that optimism: promising social transformation while deepening surveillance, reproducing bias, or ignoring community needs~\cite{berlant_cruel_2011}. These initiatives often fail due to ``design-reality gaps''~\cite{heeks_information_2002}: a mismatch between system designers' perceptions of local contexts and local actuality.

In ``tech for good'', \textit{techno-centric} projects are driven by an ``innovate at all costs'' mentality, leading to the creation of technologies with little maintenance support or local capacity building afterward~\cite{saha_towards_2022}. This mentality incentivizes a top-down design approach, where the solutions are typically designed by external technologists without adequate engagement with the local communities~\cite{walsham_ict4d_2017, dada_failure_2006, harris_how_2016, dodson_considering_2012}, especially during the decision-making stages of projects~\cite{saha_commissioning_2022}. It could also lead to the phenomenon of ``bungee jumping research''~\cite{dearden_moving_2016}, characterized by brief, superficial engagement~\cite{saha_towards_2022, sultana_parar-daktar_2019, dada_failure_2006}. Development of the solutions tends to follow a linear and short-term process: building the technology, deploying it, and moving on to the next project. Researchers often leave after their work is published, making it difficult to sustain the systems they have introduced~\cite{taylor_leaving_2013, dada_failure_2006, walsham_ict4d_2017}. Last but not least, evaluations of \textit{techno-centric} projects constrain assessments of success to technical dimensions, narrowly focusing on metrics such as model accuracy or the production of specific artifacts like software~\cite{dodson_considering_2012, lin_come_2024}.

\subsubsection{Balanced Approaches}

A \textit{balanced} approach is grounded in the understanding of problems faced by specific communities, in addition to leveraging emerging technology's capacities. Community collaborative approaches (CCA) are central to a \textit{balanced} approach. A systematic review by \citet{cooper_systematic_2022} defines CCA as projects that ``involve collaboration with community stakeholders as co-researchers throughout the research process while investigating complex problems of concern to the community (e.g., social and health-related), and developing novel technologies or sociotechnical solutions''. CCA includes work in participatory design (e.g.,~\cite{delgado_participatory_2023, bondi_envisioning_2021}), community-based participatory research (e.g.~\cite{wan_community-driven_2023, liang_surfacing_2023, strohmayer_technologies_2017}), and action research (e.g.,~\cite{hayes_relationship_2011, le_dantec_design_2016, lemaire_participatory_2011, balestrini_understanding_2014}). Projects that follow CCA are more likely to create lasting social impact because they have community support and input from the start~\cite{gandhi_digital_2007} and because they \textit{amplify} positive processes already present in the community~\cite{toyama_geek_2015}. \citet{brown_why_2019} emphasize the necessity of ``the full inclusion of all relevant parties in every aspect of the project'' to ensure project success. 

A \textit{balanced} approach emphasizes that community members' lived experiences are a legitimate form of knowledge within their specific contexts~\cite{erete_method_2023, cooper_systematic_2022}, on par with technical expertise. This emphasis is grounded in the concept of epistemic justice~\cite{fricker_epistemic_2007}, which calls for fairness in recognizing who is seen as a knower and what forms of knowledge are valued. 
In ``tech for good'' projects, this means embracing modalities of knowledge sharing that go beyond traditional academic outputs, such as storytelling and community observations~\cite{erete_method_2023, harrington_engaging_2019, zegura_care_2018}. \citet{zegura_care_2018} demonstrate that valuing the community's local knowledge is not only crucial for the quality of data and the subsequent data analysis, but also generates care among researchers and community collaborators~\cite{zegura_care_2018}. Relationships among collaborative teams are critical for the feasibility and sustainability of community-based projects~\cite{dell_ins_2016, kumar_towards_2018}. Sustainable relationships with communities can be cultivated through regular presence, trust-building, and iterative design processes, and they enable project agendas to emerge organically from (rather than being imposed on) communities~\cite{dearden_moving_2016, carroll_wild_2013}. Lastly, a \textit{balanced} approach extends evaluation beyond technical \textit{outputs} to include community \textit{outcomes} such as increased technical capacity~\cite{zegura_care_2018}, impact on everyday practice~\cite{taylor_designing_2016}, and ``empowerment of the participants and the knowledge gained by the technical designers''~\cite{drain_insights_2021}.

Prior work has started to evaluate the extent to which an AI project grants agency to community stakeholders in different stages of projects~\cite{corbett_power_2023, feffer_preference_2023, delgado_participatory_2023}. For example, \citet{delgado_participatory_2023} develop the \textit{Parameters of Participation} framework that lays out the dimensions of participatory AI work, categorizing projects into a spectrum of four modes---\textit{consult}, \textit{include}, \textit{collaborate}, and \textit{own}---and three dimensions---\textit{participation goal}, \textit{scope}, and \textit{form}. The authors then use the framework to analyze 80 research articles, revealing that existing participatory AI projects focus more on community consultation and less on collaboration and stakeholder ownership~\cite{delgado_participatory_2023}. 

Prior work has also highlighted some limitations of CCA~\cite{pine_for_2020, erete_method_2023, harrington_deconstructing_2019}. For example, nonconsensual, short-term, and non-contextual-specific participation would lead to the exploitation of impacted communities through ``participation washing''~\cite{sloane_participation_2020}. \citet{sloane_participation_2020} argue for recognition of participation as \textit{work} that is valuable for machine learning development, including data work and context-specific consultation. In AI4SG specifically, \citet{lin_come_2024} document the contributions that community organization staff make to AI4SG projects, including data work and participating in regular feedback meetings. However, this labor is often invisible or overlooked by downstream AI developers and the public~\cite{crain_invisible_2016, gray_ghost_2019, crawford_atlas_2021, sambasivan_everyone_2021}. 

\subsection{Emerging Understanding of ``Tech for Good'' Decision-Makers and AI4SG Investments}
\label{sec:rw_3}

An emerging body of work has begun to understand the perspectives, incentives, and challenges faced by grantmakers in ``tech for good.'' For example, through interviews with eight decision-makers in development projects, \citet{saha_commissioning_2022} found that community participation is rarely a part of the grant decision-making process, and it is not a requirement for applicants either. Some barriers to incorporating community voices include limited access to local organizations and communities, and overconfidence in their knowledge about the local contexts~\cite{saha_commissioning_2022}. Relatedly, \citet{gardner_ethical_2022} argue that a lack of internal expertise in AI and AI ethics creates barriers for funders to incorporate Trustworthy AI principles. Studies on other influential decision-makers, such as policymakers in government agencies, point out factors influencing their decision-making process around innovative technologies, including learning from neighboring agencies' experiences and reliance on national surveys (which omit marginalized communities' views of their needs)~\cite{saha_towards_2022}. 

Focusing specifically on private tech companies, recent work has begun to critically examine investment in AI4SG through case study analysis~\cite{nost_earth_2022, lukacz_imaginaries_2024, henriksen_googles_2022}. For example, \citet{nost_earth_2022} view Microsoft's AI for Earth program as a form of greenwashing and ``philanthrocapitalism'', where the company uses the initiative to enhance its reputation as a socially responsible corporation, while simultaneously contributing to environmental harm through its broader business operations, including its reliance on fossil fuels for data centers. 
\citet{henriksen_googles_2022} examine Google's philanthropic efforts, similarly arguing that they serve dual purposes: advancing social causes while simultaneously expanding Google’s market influence and integrating its technologies into more areas of life. Comparing ``Big Tech'' companies with the ``Big Tobacco'' industry, \citet{abdalla_grey_2021} examined the influence of ``Big Tech'' funding on the broader academic landscape (though not specifically on AI4SG), shaping events and research questions that ``put forward a socially responsible public image'' of tech companies. 

\subsection{Document Analysis of Pre-Existing Texts about AI and Funding (beyond AI)}\label{document_analysis} 
Document analysis is a systematic procedure for evaluating printed or electronic documents to ``elicit meaning, gain understanding, and develop empirical knowledge''~\cite{bowen_document_2009}. Documents are pre-existing textual instruments (e.g., public reports, proposals, meeting notes) that serve as ``social facts''~\cite{bowen_document_2009} and can be qualitatively evaluated~\cite{morgan_conducting_2022}. Document analysis can provide ``contextual richness'' by offering background information and inspiring new research questions~\cite{bowen_document_2009}. In addition, because documents are ``unaffected'' by researchers' presence and influence (unlike interview data), document analysis provides a way to evaluate stable texts that are not distorted by participant reactivity and thus offer access to information that may be difficult or impossible to gather through interactive methods~\cite{morgan_conducting_2022}. Although pre-existing texts only contain ``limited information'' and reflect only what their creators (such as funders) choose to make available, they still provide a meaningful part of the larger picture and guide future research directions~\cite{morgan_conducting_2022}.

HCI scholars have used document analysis to examine pre-existing texts about AI, encompassing AI ethics toolkits~\cite{wong_seeing_2023}, AI policy documents~\cite{wong_privacy_2023, sophie_bennani-taylor_infrastructuring_2024, bareis_talking_2022}, AI risk manifesto-style statements~\cite{oldenburg_stories_2025}, patents~\cite{cheon_powerful_2023}, and military funding~\cite{widder_basic_2024}. Pre-existing documents about AI do not simply describe AI as facts; they construct \textit{AI imaginaries}---``collective visions, beliefs, symbols, and expectations that individuals and communities hold about AI and its potential outcomes''~\cite{zhong_ai_2025}, rooted in Jasanoff and Kim's concept of \textit{sociotechnical imaginaries} (collective visions of technology's role in society)~\cite{jasanoff_containing_2009}. These documents not only define what AI is, what problems it should solve, and who benefits from it, they also ``actively shape the trajectory of AI development, integration, and governance''~\cite{zhong_ai_2025}. For example,~\citet{wong_seeing_2023} analyzed a corpus of 27 AI ethics toolkits to ``identify the discourses about ethics, the imagined users of the toolkits, and the work practices the toolkits envision and support.'' They found that AI is imagined as a technology that can be made ``ethical'' through technical adjustments---better models, checklists, or audits~\cite{wong_seeing_2023}. \citet{sophie_bennani-taylor_infrastructuring_2024} argue that policy documents perform ``discursive infrastructuring'': they stabilize certain meanings of AI and legitimize particular sociotechnical arrangements. Through an analysis of UK's National AI Strategy document, the author found that AI is constructed as 1) an autonomous, inevitable force, 2) a driver of national progress and economic modernization, and 3) a domain dominated by technical expertise over social knowledge~\cite{sophie_bennani-taylor_infrastructuring_2024}. Closest to the corpus analyzed in our work, \citet{widder_basic_2024} analyzed U.S. Department of Defense's grant solicitations for the use of AI in military applications and revealed a deep entanglement between academic AI research and militarized imaginaries, where recurring rhetorical devices justify continuous funding increases for military AI.

Similarly, prior research examining funding documents (not limited to AI) shows that funding languages are not neutral; they encode priorities, shape applicant behaviors, and structure downstream research collaborations. For example, \citet{palmero_thematic_2025} conducted a textual network analysis of Horizon Europe funding calls, treating these texts as ``cognitive artifacts of policy design'' that reveal how high-level policies are translated into operational agendas. The authors point out that funding document ``predefines applicants’ behaviour, shapes their own perceptions of eligibility, and influences evaluative processes''~\cite{palmero_thematic_2025}. Existing work in innovation studies likewise demonstrates the downstream influence of funding documents: \citet{peng_promotional_2024} find that the proportion of ``promotional language'' (such as ``innovative'' and ``unique'') in grant proposals is statistically associated with a proposal’s likelihood of being funded, while \citet{smith_understanding_2023} show ``how thematic directives rooted in funding calls influenced collaboration across research.''

\section{Methods}
To examine how funding documents frame AI4SG, we collected and analyzed funding calls and grant announcements that are publicly accessible to AI4SG project applicants and the wider public audience. In this section, we detail our approach for identifying and analyzing AI4SG funding documents (Subsections~\ref{data_dev} and~\ref{analysis}) as well as our team's positionality (Subsections~\ref{positionality}). We discuss limitations to our approach in Subsection~\ref{limitations}.

\subsection{Qualitative Dataset Development}\label{data_dev}

\paragraph{Definitions} 
In this paper, we define ``funders'' as organizations that provide financial support to AI4SG initiatives, including philanthropic foundations, nonprofit organizations, government agencies, and private technology companies. While individuals may also fund AI4SG projects, we focused on organizations because they are more likely to provide public texts about their funding agendas. We define ``funding programs'' as specific AI4SG initiatives, such as a research grant on the intersection of climate change and AI. We define ``funding documents'' as funding calls for proposals and grant announcements through which funders communicate their intentions and envisioned approaches in AI4SG. While funding calls typically outline more detailed guidelines, grant announcements present a more concise overview of an initiative’s aims, scope, and thematic priorities. We consider both as funding documents because they offer complementary insights into funding agendas. 

\paragraph{Sampling Approach}
We used the Google search engine to identify publicly available AI4SG funding documents published prior to August 2025. Following the approach in~\citet{wong_seeing_2023} when searching for AI ethics toolkits, we sought to emulate how AI4SG stakeholders (e.g., research teams, community-based organizations, individual practitioners) might seek funding opportunities. This approach also led us to discover and examine webpages from grant databases, such as \textit{nsf.gov/funding/opportunities}.

We conducted the search with keywords indicating 1) the emphasis on social good (``social good,'' ``social impact,'' ``climate change,'' ``public health,'' ``justice,'' ``equity,'' and ``community''), 2) the methodological orientation (``Artificial Intelligence,'' ``AI,'' ``Machine Learning,'' ``ML,'' or ``Data Science''), and 3) the focus on funding (``funding,'' ``fund,'' ``award,'' or ``grant''). The query returned 296 web page links. The links led to calls for proposals, announcements about the grants, blog posts about AI4SG projects, academic publications in AI4SG, academic departments and labs working on AI4SG, and so on. Using snowball sampling~\cite{naderifar_snowball_2017}, we followed references within the initial links to identify additional funding programs. For example, a university funding opportunities page\footnote{\url{https://e-hail.umich.edu/funding-opportunities/} accessed Aug 2, 2025} listed multiple AI4SG-related programs not captured in the original search. We additionally examined these funding programs for inclusion.

\paragraph{Inclusion and exclusion criteria}
We identified funding documents for inclusion using criteria reflecting three dimensions of our search terms: the emphasis on social good, the methodological orientation, and the focus on funding. The following three criteria were developed through iterative discussions between the first two authors, after jointly reviewing a random subset of 50 links from the search results. 

\begin{itemize}
    \item Funding programs must focus on addressing societal challenges such as humanitarian resource allocations, public health, and climate change. This excluded funding programs that focus on using AI for natural science discoveries (e.g., astronomical sciences~\cite{simons_foundation_nsf-simons_2023}, chemistry~\cite{nwo_new_2024}) or AI foundational and theoretical research~\cite{nsf_nsf_2024}.
    \item Funding documents must mention developing, using, or evaluating AI, Machine Learning, Data Science, or other data-driven technologies for social applications---ranging from AI as a tool to be assessed or scrutinized to AI as a proposed solution. This excluded projects on AI education (e.g.~\cite{nsf_advancing_2023}), which focus on educating people about AI instead of exploring AI as an application in the education field (e.g., AI assistants in classrooms).
    \item Funding documents must be published directly by the funders (rather than grantees), and contain details about the funding programs, including monetary funding amounts, durations of funding support, and eligibility criteria. This excluded initiatives that solely offer technical training, computing resources, or visibility without direct monetary support, such as accelerators (e.g.~\cite{apru_apru_2025}), competitions (e.g.~\cite{drivendata_competitions_2025}), and recognition awards (e.g.~\cite{maheshwari_call_2025}). Furthermore, given our interest in examining eligibility criteria across a broad spectrum of targeted applicants, we also excluded funding programs restricted to internal applicants, such as a university research seed grant available only to its faculty and students (e.g.~\cite{penn_state_university_seed_2025}). 
\end{itemize} 

Using these criteria, the first author then examined the remaining links, and the second author reviewed the selections. The first two authors then resolved any disagreements. We narrowed down to 35 funding documents (19 funding calls and 16 grant announcements) for in-depth qualitative analysis. The funding programs represent 23 funders and aggregate investments of around \$410 million. We also included supplementary materials referenced in funding documents in our qualitative dataset, as they offer important details about the funding programs, such as Frequently Asked Questions and application forms. For example, we supplemented the Google AI Impact Challenge funding call~\cite{google_google_2018} with its application guide~\cite{google_google_2019}, which laid out selection criteria and application questions. 

\paragraph{Dataset description} Table~\ref{tab:funding_doc_mr} in Appendix~\ref{app:funding_doc} provides the list of funding documents and information about the funding programs (e.g., funding amount, duration, supporting funders, target regions).\footnote{The full dataset can be accessed through this link: \\https://github.com/nicole-hjlin/chi2026-ai-for-good-funding-dataset.} Our dataset covers a wide range of programs from 2017 to 2025, aiming at both real-world implementation and use-inspired research. These funding documents were provided by government agencies (11 funding documents), philanthropic foundations (10), nonprofit organizations (7), and private technology companies (8) based in English-speaking countries. These funders were headquartered in the United States (26), the United Kingdom (5), Canada (4), Sweden (3), Germany (1), Ireland (1), and Australia (1). These funding programs targeted applications from countries including the United States (5), the United Kingdom (1), Ireland (1), and South Africa (1), as well as broader regions including globally (14), Africa (9), Latin America (2), and Low- and Medium-Income Countries (2).

\subsection{Thematic Analysis on Funding Documents}\label{analysis}
We conducted a reflexive thematic analysis~\cite{braun_using_2006, braun_reflecting_2019} of the funding documents through an iterative coding process. In the first stage of coding, the first two authors coded basic information about the funding programs, including the methods, the social issues, the duration of projects, the intended applicants, and the amount of funding. 

In the next coding stage, the first two authors independently conducted an open-ended inductive analysis of three randomly chosen funding documents and discussed the coding strategies. Through iterative discussions, the two authors identified some initial themes, which helped refine our research questions and narrow the scope of analysis to four main components of funding documents: 1) Background, 2) Expected Outcomes, 3) Eligibility Criteria and Team Composition, and 4) Funders' Support. We did not produce a fixed codebook because reflexive thematic analysis treats coding as an iterative, flexible, and interpretive process rather than a procedure of reliably categorizing data into predefined buckets~\cite{braun_reflecting_2019}. Instead, we kept detailed memos as we allowed the themes to develop inductively through ongoing, reflexive engagement with the data, researchers' memos, and discussions among the research teams.

In the third coding stage, the first author coded all the funding documents, with our research questions in mind. Each funding document's codes were reviewed by another team member. The first author then resolved any (though few) disagreements (denoted as comments on Google documents) with each code reviewer through online discussions. These discussions involved clarifying the meaning of the first author's original codes and discussing whether they aligned with the original texts in the funding documents. We did not calculate intercoder reliability, since reflexive thematic analysis treats coding as an interpretive process rather than one that seeks positivist (quantitative) reliability measures~\cite{braun_reflecting_2019}. All authors then conducted iterative affinity diagramming on an online whiteboard and discussed emerging high-level themes from the codes. The purpose of affinity diagramming was to categorize low-level codes (at the sentence level) and surface higher-level themes across the funding documents (at the section level and beyond), as well as find connections between these themes. We collaboratively characterized these high-level themes as a spectrum between a more \textit{techno-centric} and a more \textit{balanced} approach, using the lens of the prior literature synthesized in Section~\ref{sec:lens}.\footnote{Appendix~\ref{app:coding_affinity} provides screenshots of our coding process and affinity diagrams.} 

\subsection{Positionality}\label{positionality}
The authors' collective prior research and practice in public-sector AI and AI for social good have led us to believe that funding agendas are worth studying. We started with the basis that power imbalances exist within AI4SG partnerships and sought to examine power from the very beginning of AI4SG projects: when they were conceptualized for funding. While the authors are from diverse ethnic and cultural backgrounds, all are researchers in academia living in the United States and conduct research in English. 

\subsection{Limitations}\label{limitations}
The list of funding documents in our dataset is neither exhaustive nor statistically representative. Our analysis is limited to English-language materials, which excludes funding programs described in languages other than English. Additionally, funding programs that do not use the keywords in our study might have been overlooked. Finally, we emphasize that the funding documents in our corpus fall along a \textit{spectrum} from more \textit{techno-centric} to more \textit{balanced} approaches, and we caution readers against viewing this spectrum as a dichotomy. 

\section{Findings}
\label{findings}

In what follows, we report on themes in different components of funding documents and highlight rhetoric leaning towards a more \textit{techno-centric} or \textit{balanced} approach to AI4SG. We begin with the \textbf{Background} (Section~\ref{sec:finding-background}), where we discuss how funding documents framed ``social good'', ``AI'', and how AI is envisioned to help address social good problems (RQ1). In the \textbf{Expected Outcomes and Evaluations} (Section~\ref{sec:finding-expectation}), we discuss funding documents characterized outcomes, often communicated through evaluation criteria (RQ2). In the \textbf{Eligibility} (Section~\ref{sec:finding-3}) and \textbf{Support from Funders} (Section~\ref{sec:finding-approach}) sections, we shed light on \textit{how} funding documents described ``AI4SG'' can be achieved and by \textit{whom} (RQ3). 


\subsection{Background: Motivations and Beliefs Underlying AI4SG Funding Documents}
\label{sec:finding-background}

Most AI4SG funding documents in our corpus focused on addressing large-scale, complex social issues, including public health (6, e.g. [\id{C17}, \id{C19}]), development (6, e.g. [\id{A7}, \id{A8}]), climate change (3, e.g. [\id{C11}, \id{C12}]), and mental health (2, [\id{C8}, \id{C9}]). These social problems have been categorized as  ``wicked problems''---complex, large-scale, and often ill-defined problems created by multiple, interdependent, evolving issues that any single intervention cannot fully address~\cite{gurumurthy_wicked_2019, lonngren_wicked_2021}. Funding documents also rarely provided an explicit definition of what ``AI'' is.\footnote{with the exception of \id{C3} and \id{C13} (both in the Frequently Asked Questions sections)} However, some funding documents readily depicted AI-based interventions as a \textit{solution} to these large-scale social problems, emphasizing AI's transformative benefits deterministically while glossing over or omitting the specific contexts and stakeholders that AI could be beneficial for (Sections~\ref{sec:finding-411}). Some funding documents, on the other hand, were grounded in a specific context within a large-scale social issue, sometimes providing a detailed literature review. When describing AI's role in addressing the social problems, some funding documents acknowledged both the potential benefits and harms of AI in these specific contexts (Section~\ref{sec:finding-412}). 

\begin{figure}[h]
    \centering
    \includegraphics[width=\linewidth]{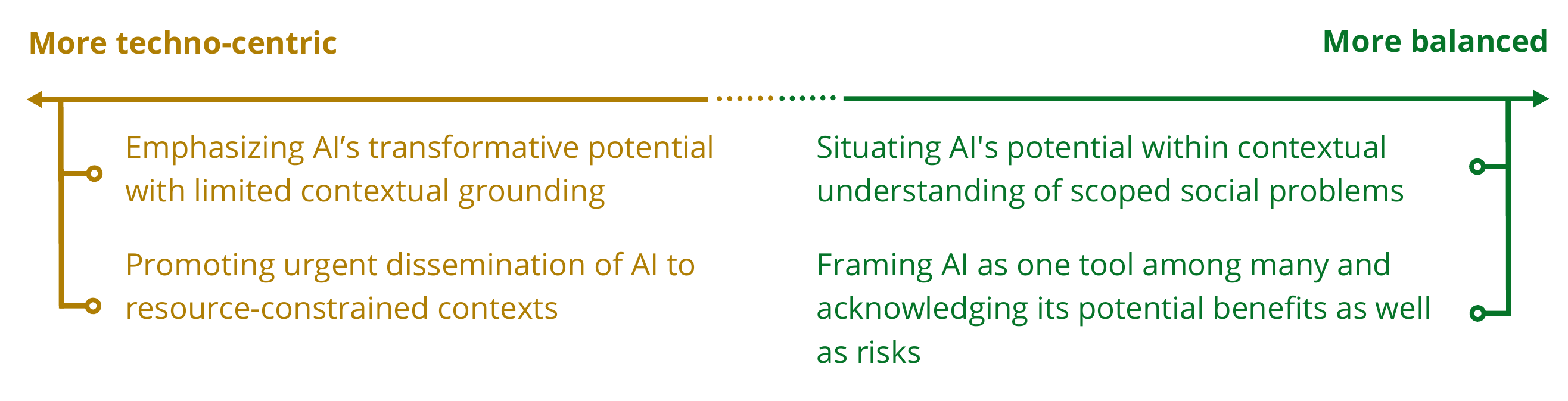}
    \caption{Themes related to rhetoric of motivations for AI4SG that lie across the spectrum between a more \textit{techno-centric} versus \textit{balanced} approach. The positions of the themes do not represent their prevalence.}
    \Description{Themes along a spectrum of more techno-centric to more balanced approaches related to motivations and backgrounds of funding documents in our corpus. More techno-centric: Emphasizing AI’s transformative potential with limited contextual grounding; Promoting urgent dissemination of AI to resource-constrained contexts. More balanced: Situating AI's potential within contextual understanding of scoped social problems; Framing AI as one tool among many and acknowledging its potential benefits as well as risks.}
    \label{fig:spec_4-1}
\end{figure}

\subsubsection{More techno-centric}
\label{sec:finding-411}
\textbf{Emphasizing AI’s transformative potential with limited contextual grounding.} Several funding documents emphasized AI's role as a potential solution to complex (and often vaguely defined) social issues, without instantiating exactly \textit{what} AI is, \textit{how} AI could be beneficial, and for \textit{whom} and what purposes. For example, UNICEF Innovation Fund Call for Data Science \& A.I. explicitly looked for \textit{``data science, machine learning, artificial intelligence or similar technology solutions''} that would broadly \textit{``benefit humanity''} (\id{C2}). The call provided example questions to elaborate on what benefiting humanity with AI means, though the application area remains broad and vague while pre-determining the use of AI as a solution: \textit{``Are you using new sources of data such as satellite imagery or social media and applying data science or artificial intelligence techniques to understand the physical world that we live in?''} Under this question, the call then further listed out examples of how AI might be beneficial, such as \textit{``[a]pplying deep learning to analyze satellite imagery and automatically map infrastructure such as schools, health centers, roads or cell towers.''} While this example provided a more specific example of how AI could be used for a specific task (using satellite imagery to automatically map infrastructure), the call did not make clear why automatically mapping infrastructure like schools might be beneficial, and for whom it would be beneficial.

Similarly, in an official announcement of Microsoft's AI for Humanitarian Action program (\id{A14}), AI was depicted as a \textit{``game changer, helping save more lives, alleviate suffering and restore human dignity by changing the way frontline relief organizations anticipate, predict and better target response efforts''} in humanitarian crises, as \textit{``[frontline relief organizations'] work by definition is often reactive and difficult to scale.''} In this depiction, AI played a transformative role in helping relief organizations \textit{``anticipate, predict and better target''} efforts. However, this depiction omitted what specific challenges relief organizations faced, and which AI could (or could not) address. 

\textbf{Promoting urgent dissemination of AI to resource constrained contexts.} Several AI4SG funding programs were designed to diffuse the benefits of AI technologies to developing countries from Western countries, and to the nonprofit sector from the private technology sector. This diffusion of AI was seen as ``urgent,'' so that under-resourced communities in developing countries or the nonprofit sector could keep up with the rapid development of AI. Some funding programs from international development agencies in Western countries, such as the AI for Development Africa program funded by IDRC (Canada) and SIDA (Sweden), emphasized the importance of ensuring \textit{``developing countries are not left behind in the AI revolution''} (\id{A9}). Some funding programs offered by private tech companies and philanthropic foundations similarly extend this narrative to the nonprofit sectors: \textit{``While private-sector businesses have been building and deploying artificial intelligence for years, most organizations in the nonprofit, civic, and public sectors have yet to robustly apply these techniques towards the complex challenges they address''} (\id{A6}). As Patrick J. McGovern Foundation's grant announcement (\id{A1}) stated, enabling nonprofits to use AI and data \textit{``have never been more urgent.''} In these funding documents, the rhetoric of AI4SG led with the assumption that AI's benefits could be directly transferred to non-Western countries and the nonprofit sectors.

\subsubsection{More balanced}\label{sec:finding-412} \textbf{Situating AI's potential within contextual understanding of scoped social problems.} Some funding programs in our corpus were motivated by not only AI's potential capacities, but also more concrete contextual needs (within a larger social domain). For example, the Call for Innovation Research Grants on Responsible Artificial Intelligence for Climate Action (\id{C11}) started with a broad focus on climate change in Africa. Within this broad focus, the call provided background information specific to the African context: \textit{``Drought is the most studied hazard across Africa. A large part of Africa is susceptible to drought, [...]. Wildfires have been more localized particularly in the recent history in Zimbabwe and Lesotho.''} The call then motivated proposals of AI use targeted towards \textit{``understanding the drivers of droughts in Africa''} and predicting drought characteristics. In the application form, the funding call then provided opportunities for applicants to specify who the proposed projects would benefit and how: \textit{``[w]ho are the target beneficiaries and what process has been undertaken to engage them in the development of the proposed action? What are the needs and challenges of each of the target groups identified and how do these relate to the action proposed?''} (\id{C11})

Another notable example is the Wellcome Mental Health Data Prize: UK and South Africa (\id{C8}), which solicited data projects that addressed a more specific problem space within mental health: \textit{``anxiety and depression in young people.''} The funding call had a regional focus on the UK and South Africa, and the funder worked with data holders from these regions to provide exemplar datasets to applicants. In addition to exemplar datasets, the funding call built on existing \textit{``evidence for different active ingredients deemed to help prevent, treat, and manage anxiety and/or depression in 14 to 24-year-olds globally.''} Applicants were then both encouraged to build on these existing findings or explore \textit{``research that expands the scope of the existing research.''} In both cases, the funding calls were motivated by not only the potential of AI, but also more specifically scoped problems within a larger social domain. 

\textbf{Framing AI as one tool among many and acknowledging its potential benefits as well as risks.} A few funding calls provided a more nuanced depiction of AI capabilities, acknowledging both its potential benefits and risks, as one part of broader problem-solving efforts. For example, the Mozilla Foundation's AI and Environmental Justice Awards (\id{C7}) began with the statement that \textit{``AI’s relationship with our current climate emergency is complex. The technology produces a vast carbon footprint, and can be used to hasten the extraction and exploitation of natural resources. Yet AI also has the potential to drive environmental justice at scale.''} The funding call (\id{C7}) also posed \textit{questions} (rather than assumptions) about whether AI may be helpful for specific use cases. Specifically, the funding call asked:
\textit{``Could AI technologies be used to catalog and analyze environmental degradation, land and water use, pollution and other environmental impacts?''} It explicitly explored AI as \textit{``a part of the solutions to these issues,''} rather than assuming that AI alone could be a solution.


\subsection{Expected Outcomes and Evaluations}
\label{sec:finding-expectation}

Funding programs in our dataset commonly aimed to produce ``innovative,'' ``scalable,'' ''feasible,'', and ``sustainable'' AI-based interventions or ``novel'' research outputs, often communicated through proposal evaluation criteria. Some funding documents emphasized the technical dimensions of desired outcomes, and some sought to showcase the benefits of AI through research outputs (Section~\ref{sec:finding-421}). In contrast, some funding documents evaluated desired outcomes beyond the technical to emphasize benefits to the community involved. Some documents also encouraged flexibility in pursuing research novelty and evaluating the negative impacts of AI technologies (Section~\ref{sec:finding-422}).

\begin{figure}[h]
    \centering
    \includegraphics[width=\linewidth]{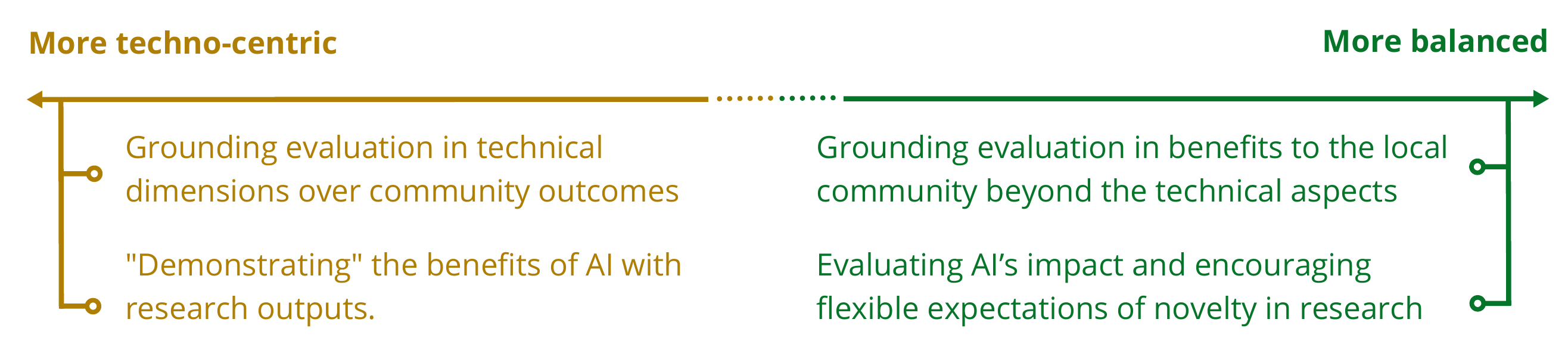}
    \caption{Themes related to funders' expected deliverables for AI4SG projects that lie across the spectrum between a more \textit{techno-centric} versus \textit{balanced} approach. The positions of the themes do not represent their prevalence.}
    \label{fig:spec_4-2}
    \Description{Themes along a spectrum of more techno-centric to more balanced approaches related to expected outcomes and evaluations of funding documents in our corpus. More techno-centric: Grounding evaluation in technical dimensions over community outcomes; “Demonstrating” the benefits of AI with research outputs. More balanced: Grounding evaluation in benefits to the local community beyond the technical aspects; Evaluating AI’s impact and encouraging flexible expectations of novelty in research.}
\end{figure}

\subsubsection{More techno-centric}
\label{sec:finding-421}
\textbf{Grounding evaluation in technical dimensions over community outcomes.} Many funding programs in our corpus aimed at deploying new or scaling existing AI systems to tackle social challenges. While funding documents rarely specified success metrics for deployed systems and who got to define success, expected outcomes were often implied through the proposal evaluation criteria. Some funding documents underscored technical dimensions of success while placing less emphasis on social processes, community empowerment, or project sustainability beyond the technology itself. For example, the Google AI Impact Challenge application guide (\id{C3}) specified criteria of successful applications, including \textit{``Impact,''} 
\textit{``Use of AI,''} \textit{``Feasibility,''} and \textit{``Scalability.''} \textit{``Impact''} asked for \textit{``a clear plan to deploy the AI model for real-world impact,''} framing societal benefit largely in relation to the effective application of AI. \textit{``Use of AI''} explicitly required applicants to integrate AI technology. While \textit{``Feasibility''} acknowledged the importance of partners and domain experts for the successful implementation, the criterion placed particular weight on the technical features of the proposed plan, asking for \textit{``a plan to access a meaningful dataset and technical expertise to apply AI to the problem.''} The \textit{``Scalability''} criterion valued projects that could expand the growth and replication of AI systems, suggesting the use of  \textit{``metrics around speed, accuracy, cost, or scalability''} in an application question. 

\textbf{``Demonstrating'' the benefits of AI with research outputs.} Other funding programs in our dataset aimed at advancing use-inspired AI research (without a requirement for real-world AI deployment), including Amazon's Data for Social Sustainability grant (\id{C6}) and Microsoft's AI for Earth program (\id{A12} and \id{A13}), both of which came from the research arms of private tech companies. Most funding documents expected the dissemination of research outputs and contributions to the research community, often in the form of \textit{``publications, presentations, code and data releases, blogs/social media posts, and other speaking engagements''} (\id{C6}), while placing limited to no emphasis on disseminating research outputs with contributors or partners in the community. Moreover, some funding programs sought to \textit{demonstrate} rather than \textit{evaluate} the uses of new AI technologies. For instance, the press announcement for the AI for Earth program (\id{A12}) described that they intended to \textit{``demonstrate how AI can deliver results more rapidly, accurately, and efficiently''} and \textit{``encourage others to innovate based on the power and potential of AI.''} This statement was grounded in the assumption that AI would be deterministically beneficial.

\subsubsection{More balanced}
\label{sec:finding-422}

\textbf{Grounding evaluation in benefits to the local community beyond the technical aspects.} A few funding programs in our corpus went beyond emphasizing technical deliverables and explicitly highlighted benefits to the community as part of their evaluation framework. For example, Lacuna Fund's call for proposals related to datasets for Natural Language Processing  (\id{C14}) and Antimicrobial Resistance (\id{C15}) specified several evaluation criteria in addition to technical quality and feasibility of proposals, particularly highlighting dimensions of \textit{``Participatory Approach''} and \textit{``Sustainability \& Communications.''} The \textit{``Participatory Approach''} criterion required applicants to \textit{``ensure sustained maintenance and usage of the dataset by the local community''} and to \textit{``share project outputs and benefits with data providers and/or the community,''} emphasizing the tangible benefits of the projects to the community involved. Similarly, the \textit{``Sustainability \& Communications''} criterion encouraged applicants to embed community-oriented structures for ongoing impact, for example, by \textit{``organizing a workshop on dataset sustainability with interested stakeholders; or establish a sustainability committee.''}

\textbf{Evaluating AI’s impact and encouraging flexible expectations of novelty in research.} Some funding calls aimed at advancing AI research explicitly supported assessments on the \textit{direction} of AI’s potential impact (rather than just the \textit{scale} of its benefits). For example, the Wellcome Mental Health Data Prize: UK and South Africa (\id{C8}) encouraged applicants to consider the risks posed by data biases in AI technologies and required applicants to address data quality, bias mitigation, and ethics in their proposals. Applicants had to consider \textit{``risks or issues associated with the proposed timelines and plans for mitigation,''} with this criterion accounting for 10\% of applicant evaluation.

Some funding calls also described expectations for research novelty more flexibly, ensuring that the exploration of AI-based interventions was suitable for the specific contexts. For example, the Wellcome Mental Health Data Prizes (\id{C8} and \id{C9}) clarified that \textit{``it is not expected that the team must innovate in their methodology if this is not appropriate for their research.''} Similarly, the Climate Change AI Innovation Grants (\id{C13}) specified that \textit{``it is perfectly acceptable to use a previously published ML algorithm or architecture,''} and grounded the exploration of AI on a solid contextual understanding: \textit{``Projects should employ or address AI/ML in a way that is well-motivated and well-scoped for the problem setting, which may or may not include the development of novel techniques.''}

\subsection{Eligibility Criteria and Team Composition}
\label{sec:finding-3}

Funding programs in our corpus commonly viewed interdisciplinary collaborations as valuable for achieving the aims of AI4SG. However, some funding calls had requirements that were biased towards (and could only be realistically met by) teams with existing AI capacities (Section~\ref{sec:finding-431}). Funding calls also differed in how they scaffolded involvement of different stakeholders, as well as who was seen as a ``domain expert'' (e.g., those with scholarly knowledge versus lived experiences). Overall, we observed that many funding documents that encouraged involvement of community stakeholders centered around consultative (rather than inclusive, collaborative, or ownership~\cite{delgado_participatory_2023}) modes of participation and only during later AI project stages (i.e., implementation phase)  (Section~\ref{sec:finding-431}). In Section~\ref{sec:finding-432}, we spotlight a few funding documents that went beyond the consultative mode and encouraged community ownership.

\begin{figure}[h]
    \centering
    \includegraphics[width=\linewidth]{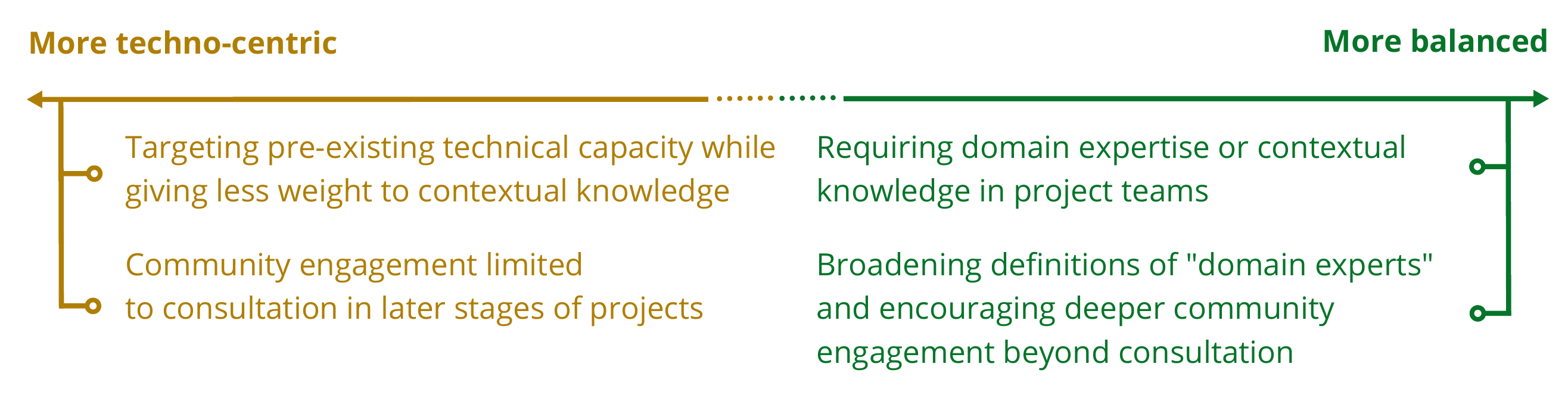}
    \caption{Themes related to edibility criteria of AI4SG funding calls that lie across the spectrum between a more \textit{techno-centric} versus \textit{balanced} approach. The positions of the themes do not represent their prevalence.}
    \Description{Themes along a spectrum of more techno-centric to more balanced approaches related to eligibility and team structures of funding documents in our corpus. More techno-centric: Targeting pre-existing technical capacity while giving less weight to contextual knowledge; Community engagement limited to consultation in later stages of projects. More balanced: Requiring domain expertise or contextual knowledge in project teams; Broadening definitions of “domain experts” and encouraging deeper community engagement beyond consultation.}
    \label{fig:spec_4-3}
\end{figure}

\subsubsection{More techno-centric}
\label{sec:finding-431}
\textbf{Targeting pre-existing technical capacity while giving less weight to contextual knowledge.} Several funding calls in our corpus designated single-applicant recipients, making equal ownership among interdisciplinary collaborations challenging. 
While most single-recipient grants in our corpus still allowed for subcontracting and encouraged partnerships, a few funding programs' eligibility criteria privileged organizations with pre-existing AI expertise, while omitting requirements for community engagement or the integration of contextual knowledge. For example, the UNICEF Innovation Fund Call for Data Science \& A.I. (\id{C2}) looked explicitly for \textit{``for-profit technology start-ups''} that had a working prototype with promising results. The intention of the grant was to \textit{``seed an exploration with a company that already exists, that has a strong team and a prototype, and that can be profitable, and successful in the space of data-science.''} While the aim for social impact was clear in the funding call, it did not mention an expectation for applicants to engage with community or domain experts to ensure adequate contextual understanding. Some grant announcements also signaled a focus on funding organizations with existing AI capacities. For instance, the Rockefeller Foundation announced a \$5.5 million investment in \textit{``satellite data \& AI to accelerate economic development and climate resilience in Africa''} (\id{A5}). The fund was granted to two organizations consisting of researchers (based in the U.S.) with \textit{``extensive experience building [...] datasets, predictive analytics models, and software platforms''} (\id{A5}).
 
\textbf{Community engagement limited
to consultation in later stages of projects.} Across many funding documents in our corpus, we observed that expectations for community engagement---in cases where they existed at all---seldom surpassed the consultation mode, where impacted communities and intended end beneficiaries could provide input, but they were not expected to shape the projects' overall scope and purpose or play a central role across all stages of the project life cycle. For example, some funding calls only encouraged community consultation during implementation, rather than in the early stages of problem identification and solution development. For example, the Google AI Impact Challenge (\id{C3}) specifically asked applicants to \textit{``identif[y] the right partners and domain experts needed for implementation.''} The funding call explained that involving these stakeholders during implementation could help \textit{``deploy the AI model [to support] real-world impact,''} where expected community engagement primarily served the purpose of AI deployment. 




\subsubsection{More balanced}
\label{sec:finding-432}

\textbf{Requiring domain expertise or contextual knowledge in project teams.} Some funding calls built interdisciplinary collaborations into team structure requirements, requiring that a variety of expertise was represented in the core project team. For example, the Artificial Intelligence (AI) for Societal Good Science Foundation Ireland Future Innovator Prize (\id{C1}) determined the structure of a core project team, comprising a \textit{``Team Lead,''} a \textit{``Team Co-Lead,''} and a \textit{``Societal Impact Champion.''} The team lead and co-lead were expected to provide technical leadership while the \textit{``societal impact champion''} was envisioned to \textit{``have appropriate experience in areas relevant to the societal impact focus of the application.''} Similarly, the Artificial Intelligence (AI) innovations in Maternal, Sexual and Reproductive Health (MSRH) call (\id{C10}) mandated that teams include a domain expert who would ensure that gender equality and inclusion were appropriately understood and considered, specifically encouraging representation from \textit{``local gender equality or women’s rights organizations or underrepresented groups.''}  

Some funding calls specifically targeted applicants from a geographic region of interest, emphasizing locally-owned projects. The Catalyzing Equitable Artificial Intelligence (AI) Use call for proposals (\id{C8}), for instance, had a specific focus on the use of Large Language Models in low- and middle-income countries (LMICs) and required that proposals were \textit{``led by investigators in LMICs.''} The funding call further clarified that: \textit{``priority will be given to proposals that demonstrate at least 80\% of the funding is going to LMIC institutions [...]''}. This requirement specified that funding would directly flow to organizations from the regions of interest. 

\textbf{Broadening definitions of ``domain experts'' and encouraging deeper community engagement beyond consultation.} A few funding calls highlighted the need to engage with community organizations or members from the affected communities directly. For example, the Wellcome Mental Health Data Prize: Africa (\id{C9}) application guide required that \textit{``individuals with lived experiences''} of mental health problems were part of the multi-disciplinary team, alongside \textit{``data scientists and mental health researchers''}. The funding call's emphasis on \textit{``integrating meaningful engagement of community and implementing partners at every stage of the research''} was built into its evaluation criteria, where engagement accounted for 10\% of an application's score. Additionally, the funding call emphasized that \textit{``contextual understanding''} and experiences in community involvement were necessary skills alongside technical proficiency, asking applicants to highlight past experiences working with communities in the application forms. 



A few funding calls required inclusion of community stakeholders as key decision-makers across the project pipeline (beyond consultation).
For example, the \textit{``Societal Impact Champion''} in project teams of the Artificial Intelligence (AI) for Societal Good Science Foundation Ireland Future Innovator Prize (\id{C1}) was expected to \textit{``identify and validate challenges in addition to advising on solution development,''} making sure that the projects were grounded in real-world needs. Beyond problem identification, the funding call also emphasized that \textit{``key measures of success [are] identified through engagement with stakeholders and beneficiaries.''} 
\subsection{Funders' Support and Connections}
\label{sec:finding-approach}

Funding calls in our dataset supported projects ranging from three months to five years (with a median of one and a half years). The longer-term programs were typically funded by governmental agencies (e.g., \id{C17} and \id{C18} by NIH, and \id{C19} by NSF). Overall, we observed a lack of provisions for post-deployment funding support. Furthermore, many funding programs readily provided non-monetary support that was biased towards technical aspects of the projects (Section~\ref{sec:finding-441}). Some other funding programs provided broader forms of support beyond the technical, sometimes through partnerships with regional organizations (Section~\ref{sec:finding-442}). However, across the funding documents in our corpus, there was a notable lack of guidance and training for applicants to develop skills in community and stakeholder engagement.

\begin{figure}[H]
    \centering
    \includegraphics[width=\linewidth]{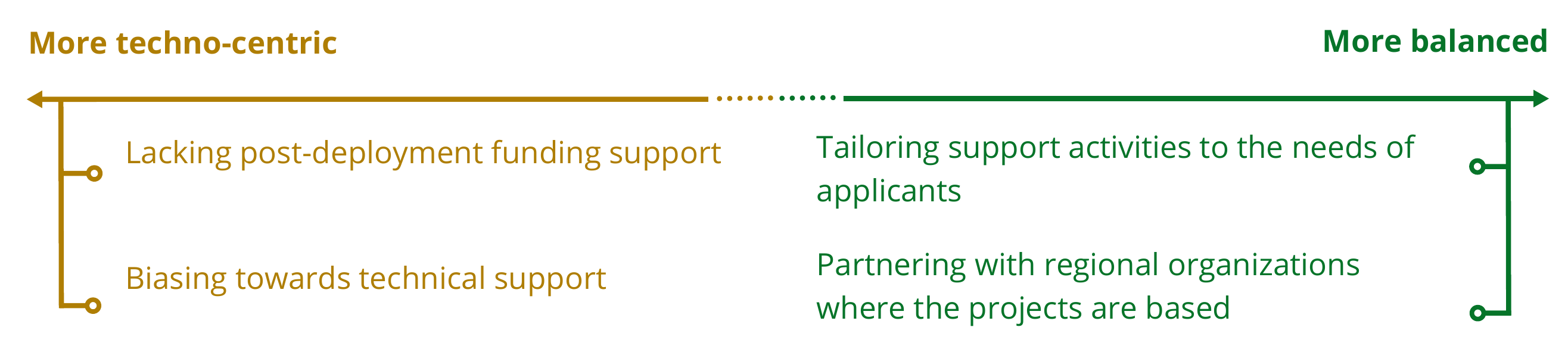}
    \caption{Themes related to the support that AI4SG funders offered that lie across the spectrum between a more \textit{techno-centric} versus \textit{balanced} approach. The positions of the themes do not represent their prevalence.}
    \Description{Themes along a spectrum of more techno-centric to more balanced approaches related to funders' support in our corpus. More techno-centric: Lacking post-deployment funding support; Biasing towards technical support. More balanced: Tailoring activities and resources to the needs of applicants beyond technical support; Partnering with regional organizations where the projects are based.}
    \label{fig:spec_4-4}
\end{figure}

\subsubsection{More techno-centric} 
\label{sec:finding-441}

\textbf{Lacking post-deployment funding support.} While sustainability was a consideration for many funding calls aiming at AI deployment, there was a notable lack of provisions for post-deployment funding support. For example, the Google AI Impact Challenge (\id{C3}) asked applicants how they would plan to \textit{``sustain and grow the impact of this work beyond this grant''} but did not clarify whether future funding would be available to support this post-deployment work. Some funding calls indicated that teams needed to sustain their deployed AI systems on their own, independent from the funders. For example,  IDRC and SIDA's AI for Climate Action Innovation call for proposals (\id{C11}) expected applicants to \textit{``source for funding for sustainability of the innovation''} themselves. 

\textbf{Biasing towards technical support.} Across the funding programs in our dataset, we also observed a notable lack of training and resources towards building contextual understanding, especially for community engagement, while technical training and access to computing resources appeared to be a standard offering. Private technology companies, in particular, often provided technical support using their own infrastructures and products. For example, Microsoft’s AI for Earth program offered \textit{``access to cloud and AI computing resources [and] technology trainings''} (\id{A12} and \id{A13}), later clarifying that this included its own \textit{``Azure compute time and our data science virtual machine offerings on Azure.''}

\subsubsection{More balanced}
\label{sec:finding-442}

\textbf{Tailoring activities and resources to the needs of applicants beyond technical support.} Some other funding programs provided broader forms of support, beyond just technical resources and training. For example, as part of the Wellcome Mental Health Data Prize (\id{C8} and \id{C9}), Wellcome Trust offered a web-based platform to facilitate knowledge-sharing across teams. The funder also facilitated matchmaking events where \textit{``individuals with different knowledge, skills and backgrounds will be introduced to collaborate and develop a proposal together''} to enable interdisciplinary collaborations. In another funding call (\id{C7}), the Mozilla Foundation described a broad array of support that funded teams would receive: \textit{``While the specific types of support are tailored to the unique needs of the awardees and cohorts, providing opportunities for peer learning and community, mentorship, and communications support are key elements of our accompaniment approach.''} However, none of the funding documents in our corpus explicitly mentioned training resources and support for local community engagement and participatory design.

A few funding documents provided references for guidelines and academic papers in responsible AI for applicants. These references could serve as a guide for applicants who were not previously familiar with these resources. For example, the funding call for AI innovations to improve Maternal, Sexual, and Reproductive Health and Rights in Sub-Saharan Africa (\id{C10}) provided a list of references on responsible AI as resources for applications. This list included~\citet{gardner_ethical_2022}'s work on the funding processes for responsible AI, which calls for the inclusion of a Trustworthy AI statement section in grant application forms. The application form included a section for responsible AI development, requiring the applicants to elaborate on \textit{``how gender equality and inclusion considerations will be integrated into the project design, implementation, monitoring and results,''} as well as considerations around \textit{``responsible AI, the collection, use and storage of data particularly individually identifiable data'' (\id{C10}).}

\textbf{Partnering with regional organizations where the projects are based.} A few funding programs involved regional partners in the design of the funding calls (\id{C8}-\id{C12}). A notable example was the design process of Wellcome Trust's Mental Health Data Prizes (\id{C8} and \id{C9}). After the first round of the data prize in the UK and South Africa (\id{C8}), the funder reflected that \textit{``it did not attract as many applications as expected from South Africa''}~\cite{wellcome_trust_request_2023}. After conducting feedback sessions~\cite{wellcome_trust_listening_2023}, the funder worked with a regional partner, the African Population and Health Research Center, to develop the funding call in Africa (\id{C9}). The regional partner played a \textit{``primary role''} in \textit{``prize management''} (e.g., developing thematic areas and leading the application selection process), \textit{``capacity building''} (e.g., providing training in data and mental health research), \textit{``matchmaking''} (e.g., reach communities and potential participants in Africa), and  \textit{``landscape analysis of existing mental health research''} (e.g., literature review on existing data sources) (\id{C9}). The partnership with the regional partner resulted in several key adjustments: providing a literature review on mental health issues in the African context to the applicants, extending the overall timeline for project development, and indicating that a \textit{``lived experience expert''} would be a member of the application selection panel.

\section{Discussion}
\label{sec:discussion}

Prior literature on ``tech for good''  establishes that to achieve desired and lasting social impacts, projects typically require a balanced emphasis on both technological capacities and contextual understanding (synthesized in Section~\ref{sec:lens}). We characterized such projects as taking a more \textit{balanced} approach. A bias towards emphasizing technological capacities and potential benefits, with limited contextual knowledge, results in a more \textit{techno-centric} approach. Using this lens to analyze funding documents, we uncovered a spectrum of conceptual framings of AI4SG and the approaches that funding rhetoric promoted. 

Our findings extend prior literature on document analysis of AI imaginaries. We show that some AI4SG funding documents in our corpus similarly reproduce many of the same \textit{techno-centric} assumptions found in ethics toolkits~\cite{wong_seeing_2023}, national AI policies~\cite{sophie_bennani-taylor_infrastructuring_2024, bareis_talking_2022}, and military funding solicitations~\cite{widder_basic_2024}, but with distinct implications for social-impact work. Like public texts that portray AI as an autonomous and inevitable force~\cite{sophie_bennani-taylor_infrastructuring_2024, bareis_talking_2022, widder_basic_2024}, some AI4SG funding calls in our corpus similarly positioned AI as a transformative solution to large-scale social problems (with limited contextual grounding) and promoted its urgent dissemination to resource-constrained social contexts (like the nonprofit sector and developing countries) (Section~\ref{sec:finding-411}). Some funding documents also mirrored AI ethics toolkits that scaffold ``technical work for individual technical practitioners''~\cite{wong_seeing_2023}, providing little to no guidance on how to involve community stakeholders despite encouraging interdisciplinary collaborations. Yet AI4SG funding documents differ from these domains in their explicit aim to advance social good, making the dissonances between expected outcomes and some \textit{techno-centric} rhetoric particularly stark. We discuss these dissonances in Section~\ref{sec:discussion1}.

At the same time, a subset of AI4SG documents in our dataset adopted more \textit{balanced} framings, such as situating AI's potential within contextual understanding of scoped social problems (Section~\ref{sec:finding-412}) and requiring domain expertise or contextual knowledge in core project teams (Section~\ref{sec:finding-432}). These examples reveal the presence of funding rhetoric that does encourage community-grounded approaches, a departure from dominant AI imaginaries. While there is complexity in how the content of funding documents influences what actually gets funded, funding rhetoric influences downstream applicants' behaviors~\cite{peng_promotional_2024} and research activities~\cite{widder_basic_2024, smith_understanding_2023}. This makes intervening in the design process of funding documents crucial for influencing downstream project approaches before projects are even selected. Building on the more \textit{balanced} approaches and existing gaps identified in our dataset, we propose opportunities for designing future funding calls for funders, including supporting relationship building, maintenance post-project deployment, and community engagement training for project teams (Section~\ref{sec:discussion2}). Finally, we outline an agenda for meaningful community engagement for funders, and specify the roles that the HCI researchers and practitioners can and should play in helping shape AI4SG funding agendas (Section~\ref{sec:discussion3})

\subsection{Revealing Dissonances Between Intended Outcomes and More \textit{Techno-centric} Approaches in AI4SG Funding Documents}
\label{sec:discussion1}

Many AI4SG funding documents in our dataset expected feasible and sustainable deployment of AI technologies or novel research advancements (Section~\ref{sec:finding-expectation}). However, the more \textit{techno-centric} approaches promoted in some funding documents may influence downstream approaches taken by AI4SG project teams, negatively impacting AI4SG's goal of achieving positive and long-lasting change in real-world contexts~\cite{perera_scaling_2024, tudor_leveraging_2024}. We highlight three prominent aspects below, in conversation with HCI literature on ``tech for good'' projects.

First of all, funding documents motivating AI's use to address large-scale social problems without instantiating concrete contextual benefits (Section~\ref{sec:finding-411}) could lead to premature emphasis on scalability, risking putting such a strong emphasis on creating something that works everywhere that it ends up not working particularly well anywhere. As \citet{hanna_against_2020} note, ``scale thinking'' necessitates homogenizing users into predefined categories, disregarding the diversity of local contexts. 
The urgency to scale could incentivize rushed implementations that are poorly suited to diverse contexts~\cite{heeks_most_2003, heeks_information_2002, dada_failure_2006}. As \citet{burrell_what_2009} put it, ``in an effort to solve problems quickly on large scale, a concern for local values, preferences, and opinions may become secondary.'' We do not argue that scalable and generalizable approaches should never be pursued, but we suggest that they should be pursued \textit{inductively}. That is, AI4SG initiatives should focus early efforts on solving specific problems of specific communities effectively and sustainably, and then look for commonalities that could inform the development of generalizable technical capabilities and infrastructures. 

Secondly, funding rhetoric that leads with assumptions of AI's deterministic benefits without acknowledging its risks (Section~\ref{sec:finding-411}) could reinforce the wider hype and ``enchanted determinism'' surrounding AI~\cite{campolo_enchanted_2020}. This narrative fosters unrealistic expectations~\cite{lin_come_2024, taylor_leaving_2013}, promotes techno-solutionism, and diverts attention from underlying social issues~\cite{walsham_ict4d_2017, karusala_womens_2017}. Furthermore, when funding documents frame AI as a force for good in the nonprofit sector and in developing countries deterministically (Section~\ref{sec:finding-411}), they risk promoting top-down approaches that cast these groups as lacking and in need of catching up~\cite{erete_method_2023, valencia_evolution_1997, dearden_moving_2016}. This framing ignores the unique needs (and assets) of the nonprofit sector and developing countries~\cite{hanna_against_2020}. While corporate donations of computing resources and support (Section~\ref{sec:finding-441}) can expand access, they may also reinforce Big Tech's hegemonic narrative~\cite{collins_black_2000, lukacz_imaginaries_2024, nost_earth_2022, seger_democratising_2023}, addressing surface-level gaps rather than deeper structural issues~\cite{ho_human-computer_2009, burrell_what_2009, dearden_moving_2016}. Notably, all the funders in our dataset are headquartered in the Global North, even though many funding programs were aimed at applicants or projects located in the Global South. This geographic asymmetry reveals an uneven distribution of hegemonic power~\cite{collins_black_2000}, as Global North institutions retain disproportionate authority to set funding agendas. These initiatives risk reproducing patterns of saviorism, positioning the Global South as a testing ground for foreign-owned systems while extracting valuable data~\cite{damilare_dosunmu_big_2025}. These projects also overlook the uneven geography of the AI supply chain; the Global South is already part of AI, but primarily through the underpaid labor that makes AI systems work~\cite{posada_coloniality_2022}.

Last but not least, funding programs that targeted applicants with existing AI capacities without scaffolding for community engagement (Section~\ref{sec:finding-431}) effectively shifted leadership power toward organizations or individuals who are technically skilled but likely lack a deep understanding of the social problem space. This emphasis also elevates AI knowledge production over community knowledge rooted in social contexts. This is fundamentally about epistemic justice---ensuring fairness in who is recognized as a ``knower'' and what forms of knowledge are legitimized~\cite{fricker_epistemic_2007}. Epistemic justice matters because when funding agendas prioritize technical capacity, they risk producing \textit{techno-centric} solutions that lack community relevance, buy-in, and sustainability~\cite{dodson_considering_2012}. 



\subsection{Aligning AI4SG Funding with Intended Outcomes: Recommendations for Funding Call Designs}
\label{sec:discussion2}

Building on the more \textit{balanced} approaches and the existing gaps we identified in our dataset, we offer the following recommendations for funding call designs, placing them in conversation with existing literature on ``tech for good'' projects. We structure these recommendations around the four key components of funding calls, to serve as a template to inform the design of future funding calls for funders, as well as future HCI research directions in community collaborative approaches.

\subsubsection{Background: situating AI4SG in well-scoped contexts} Funding calls should orient towards projects that address well-scoped and contextually relevant problems. This doesn't mean that funders should pre-define the problems, but they can support applicants by offering examples of relevant AI use cases or existing literature review on the social domain (Section~\ref{sec:finding-412}). It also involves requiring the active participation of communities during the problem formulation and solution brainstorming stage, where applicants and communities collaboratively investigate which technologies are appropriate for their specific problem. In a systematic review of community collaborative approaches, \citet{cooper_systematic_2022} highlight that many projects involve communities only during limited stages (e.g., delivery of activities), rather than throughout the full research process---including problem framing and data analysis. Yet, as \citet{bratteteig_unpacking_2016} argue, engaging communities in deciding which problems to address is key, not only for inclusion but for redistributing power and ensuring they have real influence over project direction and outcomes. This level of participation reflects a deeper form of agency than simply evaluating or implementing pre-defined solutions~\cite{delgado_participatory_2023}.

\subsubsection{Expected Outcomes: defining success with intended communities} Successful outcomes in AI4SG projects should be defined collaboratively with the communities they aim to serve. Community perspectives on success might differ from funders’ or researchers’ goals, which tend to emphasize metrics like model accuracy or academic output~\cite{lin_come_2024, erete_method_2023}. For example, \citet{lin_come_2024} found that community organizations in AI4SG partnerships value technical capacity building and project visibility more than academic output. Aligning success metrics early on helps manage expectations and reduces the risk of disappointment as projects progress~\cite{taylor_leaving_2013}. While stakeholder goals may still sometimes conflict~\cite{ho_claim_2009}, projects could aim for pre-determined mutual benefit---for instance, through fair compensation for communities' time and expertise or by supporting informal learning opportunities~\cite{carroll_wild_2013}. This would also require attention to non-traditional (and often not quantifiable) metrics that capture relationship quality~\cite{cooper_systematic_2022, erete_method_2023}. For example, funders could request letters of support and experiences from communities as part of the project's evaluation, alongside the quantitative evaluation of AI systems.

When real-world deployment is the goal, it is essential to provide funding support and maintenance resources post-deployment, which is notably absent from most funding programs in our dataset. In categorizing ICTD projects' outcomes, \citet{heeks_information_2002} calls attention to the high rate of projects that might have initially succeeded but break down over time. As \citet{jackson_rethinking_2014} argues, ``repair and maintenance'' is a critical part of technology design, yet it pervasively falls out of view when the dominant discourse of technology focuses on innovations as a production, and not a process. The maintenance work of a technology system is not only necessary for sustaining the system that will inevitably keep breaking~\cite{jackson_rethinking_2014}, it also builds relationships among workers who collaborate on the practical work~\cite{irani_critical_2014, dignazio_counting_2024}. Sustainable maintenance can involve building local capacity and intentionally transferring technology ownership to communities. For example, the research team behind \textit{SignSupport}, a mobile app co-developed with the Deaf Community of Cape Town to support pharmacy interactions, trained community members to manage a computer lab and take on system administration roles, while researchers stepped back to allow community leadership to grow without interference~\cite{dearden_moving_2016}. 

\subsubsection{Eligibility and Team Composition: requiring community co-leadership and engagement beyond consultation} Funding calls should be structured to include intended community members as co-leaders or co-applicants, ensuring they have the power to shape project direction from the outset---and importantly, that they receive funding directly. As \citet{bratteteig_unpacking_2016} point out, the ``skill composition'' of a design team significantly shapes how problems and solutions are defined. Including community members in core project roles enables them to define the problem space and success criteria, and ensures their labor and expertise are properly acknowledged and compensated~\cite{pine_for_2020}. This approach also helps redistribute power within AI4SG collaborations, supporting more equitable relationships between communities and research institutions~\cite{erete_method_2023}. Crucially, when engaging communities throughout the project, it is important to recognize and compensate for the various forms of labor they contribute---from sharing lived experience and domain expertise to facilitating connections within their communities. Future HCI research could work towards quantifying both visible and invisible labor (modeling after recent work in, for example, crowd work~\cite{toxtli_quantifying_2021} and online volunteer moderation~\cite{li_all_2022}), while also acknowledging that some forms of labor in community engagement, such as emotional labor, may be difficult or inappropriate to quantify.


\subsubsection{Funders' Support: valuing community engagement as a necessary part of AI4SG work} Funders can better support community engagement in AI4SG by recognizing that relationship-building is a foundational phase of the work, which often occurs before a clear project direction is even established~\cite{carroll_wild_2013, cooper_systematic_2022}. This phase is essential to community-collaborative approaches~\cite{cooper_systematic_2022}, and it requires the regular presence of project teams in meetings or other community contexts, which takes time and emotional labor~\cite{dearden_moving_2016}. Long-term relationships enable shared agendas to emerge, with community priorities guiding technology development from the outset~\cite{dearden_moving_2016, carroll_wild_2013}. One concrete step funders can take is to extend the typical AI4SG funding timeline (in our corpus: with a median of one and a half years) to accommodate this early-stage work. Relationship-building alone can take months~\cite{erete_method_2023} or even years~\cite{carroll_wild_2013, dearden_moving_2016}, and should be supported as a formal project phase, even prior to defining problems or solutions. One key challenge is determining appropriate timelines and support for this work. Future research in community collaborative approaches could analyze researchers’ reflections (e.g.,~\cite{erete_method_2023, dearden_moving_2016, carroll_wild_2013}) to better understand the duration and nature of relationship-building across projects. 

Furthermore, funders should also treat community engagement as a critical skill that requires training and support, on par with technical expertise. Many AI practitioners are trained in Computer Science departments that don't routinely provide training in community engagement. As \citet{okolo_you_2024} observe, AI4SG practitioners often use a hodgepodge of ad-hoc approaches to engage with end users. Funders should recognize this knowledge gap and provide clear guidance to project teams on how to properly identify stakeholders and engage with end beneficiaries. There are multiple places where this guidance could be provided: as part of the funding call itself or additional training support that many funders already provide. Funders could consider citing existing guides in community engagement from the HCI community, such as the \textit{Equity-Centered Community Design Field Guide}~\cite{creative_reaction_lab_equity-centered_2024}, the \textit{Atlanta Community Engagement Playbook}~\cite{taranji_alvarado_atlanta_2024}, and ``readiness for partnership'' checks~\cite{pine_for_2020}. Other design toolkits, like the \textit{Envisioning Cards}~\cite{friedman_envisioning_2024}, might also be useful for AI4SG project teams to identify and engage with stakeholders. Since AI introduces additional hurdles to community-based designs, including requiring intensive data work~\cite{miceli_data-production_2022, passi_making_2020, sambasivan_everyone_2021} and transparency issues~\cite{okolo_you_2024}, HCI researchers could further support funders by tailoring existing toolkits to the specific contexts of AI4SG, building on emerging Participatory AI4SG frameworks like the PACT framework~\cite{bondi_envisioning_2021}.

\subsection{Calling for Meaningful Engagement: Recommendations for the Funding Design Process}
\label{sec:discussion3}

Beyond the design of individual funding calls, we offer recommendations for funders and the HCI community on the \textit{design process} of funding programs. The design process of grantmaking is influenced by funders' connections to other stakeholders~\cite{saha_commissioning_2022}. From our dataset, we observed that private tech companies were not only providers of AI products, resources, and training support, but also funders themselves. Through investment in AI4SG, private tech companies not only expand their market influence~\cite{henriksen_googles_2022, lukacz_imaginaries_2024, nost_earth_2022} but also influence other funders' beliefs about AI's potential and their funding priorities. At the same time, the influence of communities on funding documents remains scant (c.f.~\cite{saha_commissioning_2022}). To reorient AI4SG towards a more \textit{balanced} approach, we see the need for funders to engage with intended communities and the HCI community (Section~\ref{sec:discussions_5-3-1}). We further outline implications for HCI researchers and practitioners to support the AI4SG funding design process in Section~\ref{sec:discussions_5-3-2}. 

\subsubsection{For AI4SG funders: engage with intended communities}\label{sec:discussions_5-3-1}
Engaging with communities in the design of funding calls would enable funders to align their agendas with the needs of those communities early in the decision-making process, before funding programs are even open for applications~\cite{saha_commissioning_2022, balestrini_understanding_2014}. For funding programs that target projects in a region far from the funding organizations' locations, it is essential to have regional partners with a more solid contextual understanding, such as regional research hubs and local community organizations. They can serve as ``intermediaries'' between funding organizations and intended end beneficiaries that funding organizations don't have direct access to~\cite{dell_ins_2016, saha_commissioning_2022}. 

However, several challenges may arise when funders engage with communities. \citet{gardner_ethical_2022} point out that staff in funding organizations may not have internal expertise in ethical AI development to craft and incorporate responsible AI principles into public AI funding. Relatedly, through interviews with decision-makers from grant foundations for ICTD projects, \citet{saha_commissioning_2022} found that while funders often have a desire to connect with communities, they struggle to incorporate community voices into the grantmaking process, due to barriers such as geographical distances, language barriers, and a lack of institutional incentives. Engaging with the HCI community could help address some of these challenges in incorporating community voices into the grantmaking process.

\subsubsection{For the HCI community: implications for practice and research}\label{sec:discussions_5-3-2} Our findings signal that funders are already engaging with Responsible AI guidelines and research, as evident in some funding calls that referenced academic works and guidebooks (Section~\ref{sec:finding-442}). This indicates opportunities for HCI research and practical tools to inform funders' design process. Following up on Saha et al.'s call on HCI researchers and practitioners to serve as ``mediators'' between decision-makers and communities' voices~\cite{saha_towards_2022}, we see opportunities in the AI4SG space where the HCI community can mediate between communities, funders, and other stakeholders like funders' regional partners. 

\paragraph{For HCI practitioners} While there is a rich body of resources for community collaborative approaches~\cite{cooper_systematic_2022}, there exist gaps for the HCI community to bridge between frameworks and practices, such as implementing formal evaluations of participatory design projects~\cite{bossen_evaluation_2016, nguyen_evaluation_2022}. One possible reason for these gaps is the lack of HCI expertise present in non-academic institutions. HCI practitioners, who work directly with non-academic stakeholders to implement frameworks, can establish partnerships with or apply for positions at funding organizations to put community collaborative approaches into practice (such as consulting contracts or sabbatical positions at funding organizations like the National Science Foundation in the U.S.). 

Specifically, HCI practitioners can support funders in identifying key stakeholders to engage with in communities, designing training materials on community engagement for project teams, and providing community engagement workshops for both the project teams and the funders. Moreover, HCI practitioners can help foster reflexivity in the grantmaking process, such as reflection workshops that bring funders, regional partners, representatives of intended beneficiaries, and targeted applicants together to learn from past funding program experiences, leveraging deliberation tools when considering strategies around AI (such as the Situated AI Guidebook~\cite{kawakami_situate_2024}). Wellcome Trust's design process of Mental Health Data Prizes (\id{C8} and \id{C9}) exemplifies the benefits of feedback workshops (Section~\ref{sec:finding-442}).  

\paragraph{For HCI researchers}
Our analyses also surface opportunities for future HCI research to further understand the role of funders in AI4SG initiatives, as well as pathways to shifting towards more \textit{balanced} practices. We list a few promising research directions below: 
\begin{itemize}
    \item One promising direction is to interview staff members of AI4SG funding organizations to better understand the motivations, internal deliberations, and tensions between what is articulated in funding documents and the project selection process. Our document analysis can serve as a basis for developing interview protocols for funders, potential applicants, and funded teams~\cite{bowen_document_2009}.
    \item Future work could also engage with funded project teams to better understand their funding application strategies and needs (c.f.,~\cite{foster_evaluating_2024,baxter_evaluating_2016}).
    \item Linking funding documents to funded project outcomes could be fruitful for identifying alignments or gaps between the visions communicated through funding rhetoric and practices on the ground.
    \item Future work could also categorize commonalities of success and failure specific to AI4SG initiatives (similar to \citet{dodson_considering_2012}'s analysis of ICTD projects). Since publicly available written materials may be biased toward success stories, focusing on one or a few case studies could yield insightful analysis.
    \item Future work could expand the analysis to include AI4SG funding documents published in languages other than English to identify funders outside Western societies and examine their funding agendas.
\end{itemize}

Furthermore, the HCI research community can offer counter-narratives to the more \textit{techno-centric} imaginaries of AI in funding calls, such as those that emphasize diverse ways of knowing and sustainable development of AI within planetary boundaries. Some emerging critical reassessments of current AI design practices and public discourse around AI include work on AI abandonment~\cite{johnson_fall_2024}, post-capitalist AI systems based on feminist values~\cite{varon_ai_2024}, Indigenous thoughts~\cite{ofosu-asare_cognitive_2024, lewis_abundant_2024}, post-colonial computing~\cite{sultana_witchcraft_2019}, and post-growth philosophies~\cite{sharma_post-growth_2023, kuijer_designing_2024}. HCI scholars can make these perspectives more accessible to funders and AI4SG stakeholders by translating them into public-facing texts, such as articles in ACM Interactions magazine and online posts. These textual instruments could help bridge the gap between academic insights and real-world decision-making.

\section{Conclusion}

Artificial Intelligence for Social Good (AI4SG) emerges as a growing body of research and practice that explores the potential of AI and data-driven technologies to tackle social problems like public health~\cite{shi_artificial_2020, perrault_ai_2020, tomasev_ai_2020, rolnick_tackling_2022}. A growing body of work shows that AI4SG initiatives face various challenges in real-world contexts, including deployment and project sustainability~\cite{lin_come_2024, aula_stepping_2023, ben_green_good_2019}. While prior work has focused on the roles that \textit{project teams}---typically AI developers and their nonprofit partners--play in project outcomes, an emerging body of work points to the significant influence of \textit{funding agendas} in shaping the direction, design, and priorities of technology initiatives aimed at social impact~\cite{lin_come_2024, saha_commissioning_2022, gardner_ethical_2022}.

In this study, we begin addressing this gap and ``study up'' AI4SG funding documents, examining them through the lens of existing literature on ``tech for good'' projects. Our analysis reveals dissonances between AI4SG funders’ intentions for positive, sustainable social impact and the \textit{techno-centric} approaches that some funding programs promoted. At the same time, we identified several more \textit{balanced} approaches as well as existing gaps in funding calls. Building on our analysis, we offer recommendations for funders on funding call designs, while emphasizing the importance of community engagement in the funding call design process. Finally, we specify the roles that HCI practitioners and researchers can play to mediate between AI4SG stakeholders, supporting meaningful engagement with communities in the early stages of AI4SG projects---when funding programs are being designed.

\begin{acks}
We are grateful to the anonymous reviewers, Weiwei Pan, Eura Shin, and members of the \textit{Intelligent Interactive Systems} Group at Harvard and the \textit{Data + Feminism Lab} at MIT for their valuable feedback on the early drafts of this work.
\end{acks}

\bibliographystyle{ACM-Reference-Format}
\bibliography{references}

\appendix
\section{Funding documents}
\label{app:funding_doc}
See Table~\ref{tab:funding_doc_mr} next page. The full dataset can be accessed through this link: \\https://github.com/nicole-hjlin/chi2026-ai-for-good-funding-dataset.

\begin{landscape}
\onecolumn
\renewcommand{\arraystretch}{1.2}
\scriptsize
\begin{longtable}[H]{|p{0.3cm}|p{2.6cm}|p{0.5cm}|p{1.5cm}|p{0.9cm}|p{1.2cm}|p{1cm}|p{0.6cm}|p{1.2cm}|p{1.6cm}|p{1.8cm}|p{0.5cm}|}
\caption{Funding Documents Related to AI and Social Good} \label{tab:funding_doc_mr} \\
\hline
\textbf{ID} & \textbf{Title} & \textbf{Year} & \textbf{Funding Amount} & \textbf{Funding Duration} & \textbf{Funder(s)} & \textbf{Funder Types} & \textbf{Funder HQ} & \textbf{Eligible Regions} & \textbf{Social Domains} & \textbf{Eligible Applicants} & \textbf{Ref.} \\ \hline
\endfirsthead
\hline
\textbf{ID} & \textbf{Title} & \textbf{Year} & \textbf{Funding Amount} & \textbf{Funding Duration} & \textbf{Funder(s)} & \textbf{Funder Types} & \textbf{Funder HQ} & \textbf{Eligible Regions} & \textbf{Social Domains} & \textbf{Eligible Applicants} & \textbf{Ref.} \\ \hline
\endhead
\hline \multicolumn{12}{|r|}{\textit{Continued on next page}} \\ \hline
\endfoot
\hline
\endlastfoot

\idanchor{C1} & Artificial Intelligence (AI) for Societal Good Science Foundation Ireland Future Innovator Prize & 2019 & multi-phased (Concept: €20K; Seed: €200K; Prize: €2M) & multi-phased (Concept: 3 months; Seed: 9 months; Prize: 12 months) & Science Foundation Ireland & Government Agency & Ireland & Ireland & UN SDGs & ``Academic Staff or Contract Researcher; Holds a PhD (or equivalent)'' & \cite{science_foundation_ireland_artificial_2019} \\ \hline
\idanchor{C2} & UNICEF Innovation Fund Call for Data Science \& A.I & 2019 & \$100{,}000 per project & not specified & UNICEF & Non-profit & United States & Global & broadly ``benefit humanity'' & ``private company in a UNICEF programme country'' & \cite{unicef_unicef_2019} \\ \hline
\idanchor{C3} & Google AI Impact Challenge (Continued as AI for the Global Goals) & 2018 & \$25 million in total (for 20 organizations) & 1--3 years & Google & Private Tech Company & United States & Global & broadly ``social and environmental challenges'' & ``Organizations that are experienced in AI to those with an idea for how they could put their data to better use'' & \cite{google_google_2018} \\ \hline
\idanchor{C4} & A Grand Challenges Request for Proposals: Catalyzing Equitable Artificial Intelligence (AI) Use & 2023 & \$100{,}000 per project (\$3{,}000{,}000 total) & up to 3 months & Bill \& Melinda Gates Foundation & Philanthropy Foundation & United States & Low- and Medium-Income Countries (LMICs) & Decision-making in LMICs & ``investigators in LMICs'' & \cite{the_bill__melinda_gates_foundation_grand_2023} \\ \hline
\idanchor{C5} & Innovative Data and Modeling Approaches to Measure Women's Health & 2025 & Up to \$150{,}000 per project & 18--24 months & Bill \& Melinda Gates Foundation & Philanthropy Foundation & United States & Global (preferring Sub-Saharan Africa and South Asia) & Women's Health & ``nonprofit organizations, for-profit companies, international organizations, government agencies and academic institutions'' & \cite{bill__melinda_gates_foundation_innovative_2025} \\ \hline
\idanchor{C6} & Data for Social Sustainability call for proposals & 2021 & $le{}$ \$80{,}000 (avg) + AWS credits ($\le{}$ \$20{,}000 avg) & 1 year & Amazon & Private Tech Company & United States & Global & human rights protections & ``full-time faculty member or permanent NGO researcher'' & \cite{amazon_data_2021} \\ \hline
\idanchor{C7} & Mozilla Technology Fund: AI and Environmental Justice & 2024 & \$50{,}000 per project (\$300{,}000 total) & 1 year & Mozilla Foundation & Non-profit & United States & Global & environmental justice & ``all applicants regardless of geographic location or institutional affiliation'' & \cite{mozilla_foundation_mozilla_2023} \\ \hline
\idanchor{C8} & Wellcome Mental Health Data Prize: UK and South Africa & 2022 & multi-phased (Discovery: £40{,}000; Prototyping: £100{,}000; Sustainability: £500{,}000) & multi-phased (each phase 6 months) & Wellcome Trust & Philanthropy Foundation & United Kingdom & United Kingdom and South Africa & anxiety and depression in young people & ``higher education, research institutes, non-academic healthcare orgs, not-for-profits'' & \cite{wellcome_trust_wellcome_2022} \\ \hline
\idanchor{C9} & Wellcome Mental Health Data Prize: Africa & 2024 & £200{,}000 per team (10 teams total) & 1 year & Wellcome Trust & Philanthropy Foundation & United Kingdom & Africa & anxiety, depression, and psychosis & ``lead applicants based in Africa (academic, non-profits on mental health, government agencies, research institutions)'' & \cite{aphrc_mental_2024} \\ \hline
\idanchor{C10} & Request for Applications (RFA) for AI innovations in MSRH & 2022 & \$10{,}000--\$15{,}000 (students/startups); \$30{,}000--\$40{,}000 (established orgs) & 1.5 years & IDRC and Sida & Government Agency & Canada and Sweden & Sub-Saharan Africa & maternal, sexual, and reproductive health/rights & ``students and start-up organisations; established organisations'' & \cite{hash_request_2022} \\ \hline
\idanchor{C11} & Call for proposals: AI for climate action Innovation Research projects & 2022 & \$50{,}000--\$60{,}000 & 1.5 years & IDRC and Sida & Government Agency & Canada and Sweden & Africa & climate change & ``researchers from accredited universities; public/private research institutions'' & \cite{ruforum_call_2022} \\ \hline
\idanchor{C12} & Innovate Africa Challenge Artificial Intelligence (AI) for Climate Action & 2024 & \$50{,}000 & multi-phased (bootcamp; incubation; implementation) & BMZ & Government Agency & Germany & Africa & climate change & company from any of the 39 Smart Africa Alliance countries & \cite{smart_africa_about_2024} \\ \hline
\idanchor{C13} & Call for Proposals: Climate Change AI Innovation Grants & 2024 & up to \$150{,}000 per proposal (up to \$1.4M total) & 1 year & Climate Change AI (QCF, Google DeepMind, Global Methane Hub) & Non-profit & United States & Global & climate change & PI affiliated with accredited university in one of 38 OECD member countries & \cite{climate_change_ai_climate_2024} \\ \hline
\idanchor{C14} & Request for Proposals: Natural Language Processing (NLP) & 2024 & \$1{,}000{,}000 & 1.5 years & Lacuna Fund (supported by Google.org) & Non-profit & United States & Africa and Latin America & NLP data & ``non-profit entity, research institution, for-profit social enterprise, or team of such orgs'' & \cite{lacuna_fund_request_2024-1} \\ \hline
\idanchor{C15} & Request for Proposals: Antimicrobial Resistance (AMR) & 2024 & \$1{,}000{,}000 & 1.5 years & Lacuna Fund (supported by Wellcome Trust) & Non-profit & United Kingdom & LMICs in Africa, Latin America, South \& Southeast Asia & public health & ``non-profit entity, research institution, for-profit social enterprise, or team of such orgs'' & \cite{lacuna_fund_request_2024} \\ \hline
\idanchor{C16} & Public-Private Partnerships to Improve Population Health Using AI/ML & 2025 & Up to \$612{,}500 per project (Planning + Implementation) & 18 months & NIH & Government Agency & United States & United States & public health & ``Public-Private dyad or triad applicant teams'' & \cite{aim-ahead_aim-ahead_2025} \\ \hline
\idanchor{C17} & Leveraging Big Data Science to Elucidate the Mechanisms of HIV Activity and Interaction with Substance Use Disorder & 2023 & \$2{,}000{,}000 & Up to 5 years & NIH & Government Agency & United States & Global & public health & Organizations & \cite{national_institutes_of_health_nih_expired_2025} \\ \hline
\idanchor{C18} & DS-I Africa Research Hubs & 2020 & \$32{,}500{,}000 & Up to 5 years & NIH & Government Agency & United States & Africa & public health & ``Non-domestic (non-U.S.) Entities (Foreign Institutions)'' & \cite{national_institutes_of_health_nih_expired_2020} \\ \hline
\idanchor{C19} & Smart Health \& Biomedical Research in the Era of AI \& Advanced Data Science (SCH) & 2025 & Up to \$20{,}000{,}000 per year & Up to 4 years & NSF & Government Agency & United States & United States & public health & Institutions of Higher Education; Non-profit, non-academic organizations & \cite{us_national_science_foundation_nsf_2025} \\ \hline

\idanchor{A1} & ``[Patrick J. McGovern] Foundation commits \$40M to advancing AI and data for good'' & 2021 & \$40{,}000{,}000 & not specified & Patrick J. McGovern Foundation & Philanthropy Foundation & United States & unspecified & broadly nonprofits & Nonprofits and NGOs; orgs on data dignity, data stewardship, AI ethics & \cite{patrick_j_mcgovern_foundation_foundation_2021} \\ \hline
\idanchor{A2} & ``Mastercard Center for Inclusive Growth and data.org launch Artificial Intelligence to Accelerate Inclusion challenge'' & 2024 & not specified & not specified & Mastercard Center for Inclusive Growth; data.org & Non-profit & United States & Global & economic empowerment & ``the diverse range of communities often left out of AI for social impact'' & \cite{mastercard_mastercard_2024} \\ \hline
\idanchor{A3} & ``DataKind to help more organizations use data for social change with new \$1.1 million investment from Knight Foundation'' & 2015 & \$1{,}100{,}000 & 3 years & John S. and James L. Knight Foundation & Philanthropy Foundation & United States & Global & broadly nonprofits & DataKind & \cite{knight_foundation_datakind_2015} \\ \hline
\idanchor{A4} & ``AI for local news: advancing business sustainability in newsrooms'' & 2021 & \$3{,}000{,}000 & 2 years & John S. and James L. Knight Foundation & Philanthropy Foundation & United States & United States & local journalism & ``local news organizations'' & \cite{knight_foundation_ai_2021} \\ \hline
\idanchor{A5} & ``The Rockefeller Foundation Invests in Satellite Data \& AI to Accelerate Economic Development and Climate Resilience in Africa'' & 2022 & \$5{,}500{,}000 & 3 years & The Rockefeller Foundation & Philanthropy Foundation & United States & United States & development & e-GUIDE and Atlas AI & \cite{the_rockefeller_foundation_rockefeller_2022-1} \\ \hline
\idanchor{A6} & ``The Rockefeller Foundation Commits \$1M to Expand BlueConduit’s AI-Based Solution to America’s Lead Pipe Challenge'' & 2022 & \$1{,}000{,}000 & 5 years & The Rockefeller Foundation & Philanthropy Foundation & United States & United States & infrastructure & BlueConduit & \cite{the_rockefeller_foundation_rockefeller_2022} \\ \hline
\idanchor{A7} & ``Combining forces for a new phase of AI for development: Africa and beyond'' & 2023 & CAD130{,}000{,}000 & 5 years & IDRC; UK FCDO; Bill \& Melinda Gates Foundation; USAID & Government Agency, Philanthropy Foundation & Canada; United Kingdom; United States & Africa & development & unspecified & \cite{idrc_combining_2023} \\ \hline
\idanchor{A8} & ``Artificial Intelligence for Development (AI4D)'' & 2020 & CAD 20{,}000{,}000 & 5 years & IDRC and Sida & Government Agency & Canada and Sweden & Africa & development & unspecified & \cite{idrc_artificial_2020} \\ \hline
\idanchor{A9} & ``UK unites with global partners to accelerate development using AI'' & 2023 & £38{,}000{,}000 & 5 years & the UK government & Government Agency & United Kingdom & Africa & development & unspecified & \cite{uk_government_uk_2023} \\ \hline
\idanchor{A10} & ``data.org Announces Generative AI Skills Challenge Awardees'' & 2023 & \$250{,}000 per organization (5 orgs) & not specified & data.org; Microsoft & Private Tech Company & United States & Global & economic empowerment & Organizations broadly & \cite{dataorg_dataorg_2023} \\ \hline
\idanchor{A11} & ``Inclusive Growth and Recovery Challenge'' & 2021 & \$10{,}000{,}000 & not specified & data.org; Mastercard Center for Inclusive Growth; The Rockefeller Foundation; The Paul Ramsay Foundation & Non-profit & United States; Australia & unspecified & development & unspecified & \cite{dataorg_inclusive_2021} \\ \hline
\idanchor{A12} & ``Announcing AI for Earth: Microsoft’s new program to put AI to work for the future of our planet'' & 2017 & \$2{,}000{,}000 & 1 year & Microsoft & Private Tech Company & United States & Global & sustainability & ``researchers and organizations'' & \cite{smith_announcing_2017} \\ \hline
\idanchor{A13} & ``AI for Earth can be a game-changer for our planet'' & 2017 & \$50{,}000{,}000 & 5 years & Microsoft & Private Tech Company & United States & Global & sustainability & ``universities, nongovernmental organizations and others'' & \cite{smith_ai_2017} \\ \hline
\idanchor{A14} & ``Using AI to help save lives'' & 2018 & \$40{,}000{,}000 & 5 years & Microsoft & Private Tech Company & United States & Global & humanitarian crisis & ``nongovernmental organizations (NGOs) and humanitarian organizations'' & \cite{smith_using_2018} \\ \hline
\idanchor{A15} & ``Google.org Impact Challenge: Tech for Social Good'' & 2023 & €16{,}000{,}000 & up to 6 months & Google & Private Tech Company & United States & Europe & sustainability, economic opportunity, cybersecurity & ``European organisations with projects focused on sustainability, economic opportunity or cyber security'' & \cite{google_googleorg_2023} \\ \hline
\idanchor{A16} & ``Google.org Accelerator: Generative AI open call'' & 2025 & \$30{,}000{,}000 & 6 months & Google.org & Private Tech Company & United States & Global & broadly nonprofits & social impact organizations & \cite{googleorg_googleorg_2025} \\ \hline

\end{longtable}
\end{landscape}

\section{Coding Process and Affinity Diagram}
\label{app:coding_affinity}

\begin{figure}[H]
    \centering
\includegraphics[width=0.85\linewidth]{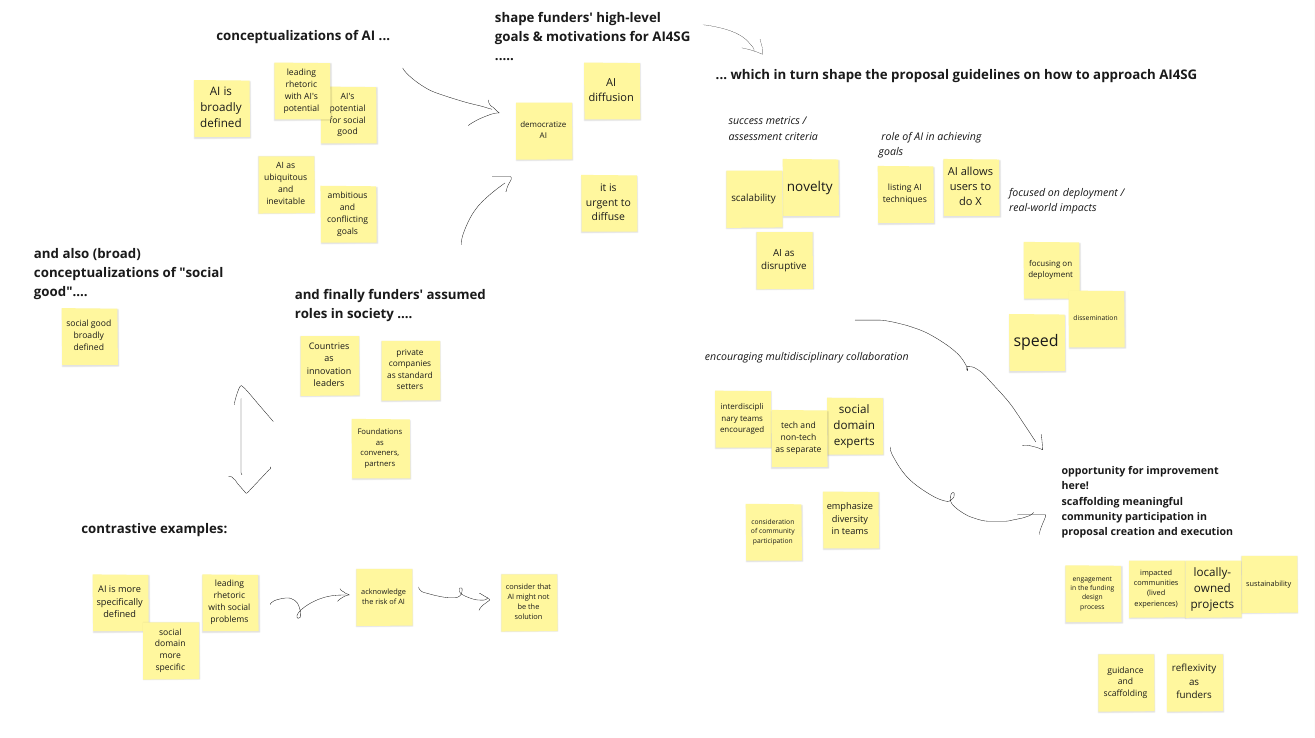}
    \caption{The first two authors independently conducted an open-ended inductive qualitative analysis of three randomly chosen funding documents. Through iterative discussions and collective affinity diagramming on an online whiteboard, the two authors identified some initial themes, which helped refine our research questions and coding strategies.}
    \Description{This figure shows a screenshot of an online whiteboard where sticky notes of emerging themes are grouped into clusters. Arrows are used to link clusters and indicate relationships between them.}
    \label{fig:aff_diag_1}
\end{figure}

\begin{figure}[H]
    \centering
    \includegraphics[width=1\linewidth]{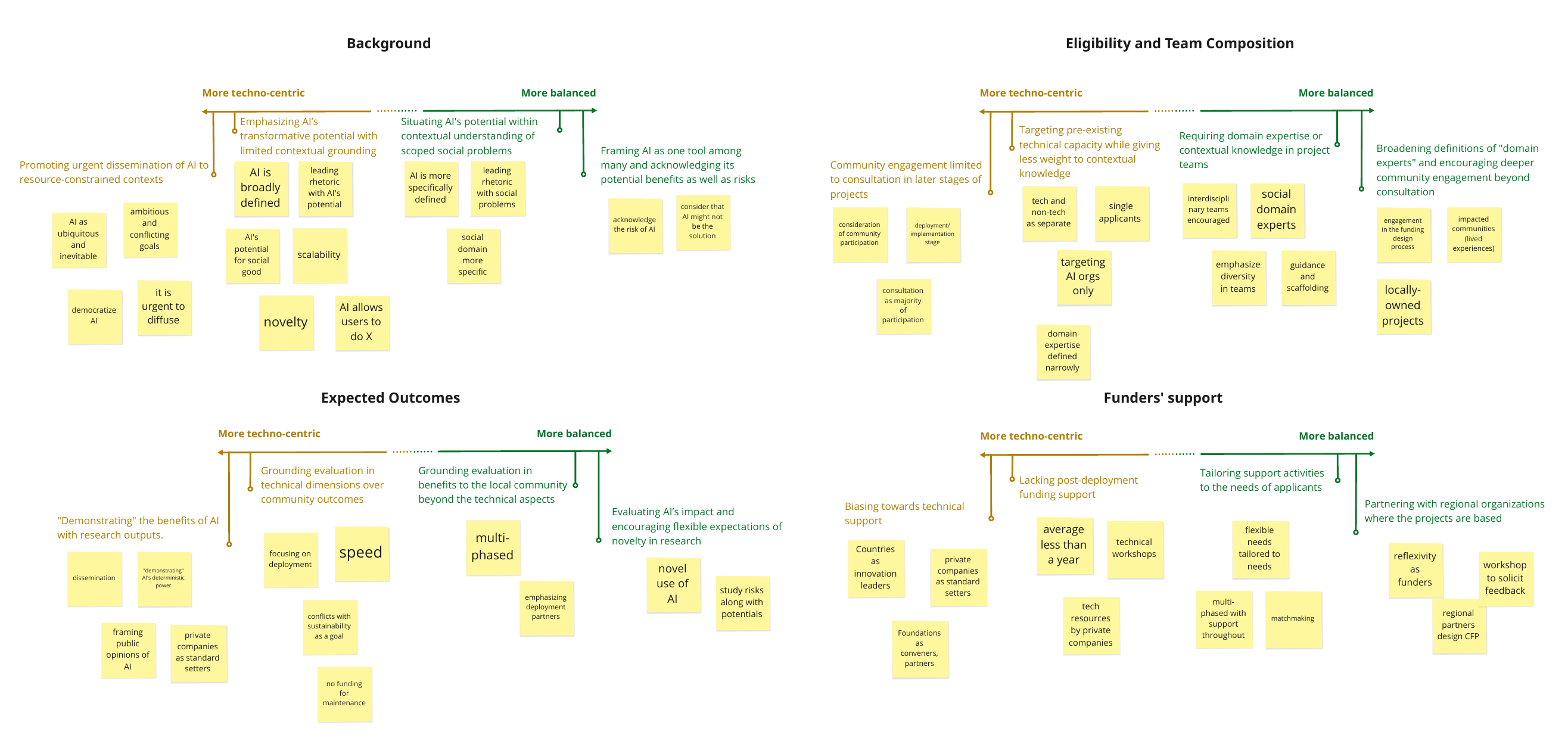}
    \caption{After an open-ended inductive coding process (Figure~\ref{fig:aff_diag_1}), we narrowed the scope of analysis to four main components of funding documents: 1) Background, 2) Expected Outcomes, 3) Eligibility Criteria and Team Composition, and 4) Funders' Support. We characterized high-level themes into a spectrum between a more \textit{techno-centric} approach or a more \textit{balanced} approach through collective affinity diagramming.}
    \Description{This figure shows a screenshot of an online whiteboard where sticky notes of codes are grouped into clusters under different themes. These themes are placed along a spectrum of more balanced and more techno-centric under each funding document component: background (on the top left corner), expected outcomes (on the bottom left corner), eligibility and team composition (on the top right corner), and funders' support (on the bottom right corner).}
    \label{fig:aff_diag_2}
\end{figure}

\end{document}